\documentclass[12pt,twoside,english]{article}
\usepackage[T1]{fontenc}
\usepackage[latin1]{inputenc}
\usepackage{geometry}
\usepackage{babel}
\usepackage{graphics}


\begin{document}

\begin{titlepage}

\setcounter{page}{1}
\rightline{}
\vfill
\begin{flushright}
UFIFT-HEP-03-1
\end{flushright}

\begin{center}
 {\Large \bf The Kinetic Interpretation of the DGLAP Equation, \\
its Kramers-Moyal Expansion\\
and Positivity of Helicity Distributions \\ }

\vfill
\vfill
 {\large Alessandro Cafarella$^{(1)}$ and
 Claudio Corian\'{o}$^{(1)\, (2)}$
}

\vspace{.12in}
 {\it  $^{(1)}$Dipartimento di Fisica, Universita' di Lecce \\
 and INFN Sezione di Lecce \\ Via Arnesano 73100 Lecce, Italy \\
 and \\
 $^{(2)}$Institute for Fundamental Theory\\
 University of Florida \\
 Gainesville, FL, 32611\\ 
 alessandro.cafarella@le.infn.it, claudio.coriano@le.infn.it} 

\vspace{.075in}

\end{center}
\vfill
\centerline{ \em Dedicated to Prof. Pierre Ramond for his 60th birthday}

\begin{abstract}
According to a rederivation - due to Collins and Qiu - 
the DGLAP equation 
can be reinterpreted (in leading order) in a probabilistic way.
This form of the equation has been used indirectly to prove the bound 
$|\Delta f(x,Q)| < f(x,Q)$ between polarized and unpolarized distributions, or positivity 
of the helicity distributions, for any $Q$. 
We reanalize this issue by performing a detailed numerical study of the positivity bounds of the helicity distributions. 
To obtain the numerical solution
we implement an x-space based algorithm for polarized and 
unpolarized distributions
to next-to-leading order in $\alpha_s$, which we illustrate.  
We also elaborate on some of the formal properties of the Collins-Qiu form 
and comment on the underlying regularization, 
introduce a Kramers-Moyal expansion of the equation and
briefly analize its Fokker-Planck approximation. 
These follow quite naturally
once the master version is given. We illustrate this expansion both for the valence 
quark distribution $q_V$ and for the transverse spin distribution $h_1$.

\end{abstract}
\smallskip

\end{titlepage}

\setcounter{footnote}{0}

\def\beq{\begin{equation}}
\def\eeq{\end{equation}}
\def\beqn{\begin{eqnarray}}
\def\eeqn{\end{eqnarray}}
\def\ie{{\it i.e.}}
\def\eg{{\it e.g.}}
\def\half{{\textstyle{1\over 2}}}
\def\nicefrac#1#2{\hbox{${#1\over #2}$}}
\def\third{{\textstyle {1\over3}}}
\def\quarter{{\textstyle {1\over4}}}
\def\m{{\tt -}}

\def\p{{\tt +}}

\def\slash#1{#1\hskip-6pt/\hskip6pt}
\def\slk{\slash{k}}
\def\GeV{\,{\rm GeV}}
\def\TeV{\,{\rm TeV}}
\def\y{\,{\rm y}}

\def\l{\langle}
\def\r{\rangle}

\setcounter{footnote}{0}
\newcommand{\beqa}{\begin{eqnarray}}
\newcommand{\eeqa}{\end{eqnarray}}
\newcommand{\eps}{\epsilon}

\pagestyle{plain}
\setcounter{page}{1}

\section{Introduction}
QCD, the theory of the strong interactions,
has reached a stage in which precision measurements of its dynamics
have become possible.
This scrutiny has allowed
to obtain a better understanding of the fundamental
structure of the nucleons, disentangling some important features of the
underlying quark-gluon interaction with a very good accuracy.

A lot of effort has been undertaken in the
last few years to extend the same picture 
- at the same level of accuracy - also to polarized collisions, with a systematic perturbative
analysis performed up to next-to-leading order (NLO) and, in part, in the unpolarized case, also to
next-to-next to leading (NNLO)
in $\alpha_s$, the QCD coupling constant.

The aim of this theoretical and experimental effort, in particular at RHIC, 
the Relativistic Heavy Ion Collider at Brookhaven, is to describe 
the polarized spin distributions of the nucleon with accuracy. 

Therefore the study of possible theoretical constraints on the form 
of the initial conditions for these 
distributions and their evolution under the renormalization group (RG) 
turns out to be very useful.

An interesting constraint relating longitudinally polarized, unpolarized and transversely polarized distributions is Soffer's inequality, which deserves a special attention, since has to be
respected by the evolution to any order in $\alpha_s$. Some tests of the
inequality have been performed in the near past, bringing support to it. 
However, other inequalities are supposed to hold as well. 

In this work we perform a NLO analysis of an inequality which relates longitudinally polarized distributions and unpolarized ones. The inequality can be summarized in the statement 
that helicity distributions (positive and negative) for quarks and gluons have to be 
positive. The inequality states that

\beq
|\Delta f(x,Q^2)| < f(x,Q^2) 
\eeq
or
\beq
f^\pm (x,Q^2) > 0  
\label{helicities}
\eeq
where the $\pm$ refers to the the possible values of the helicities 
of quarks and gluons. The statement is supposed to hold, at least in 
leading order, for any $Q$. To analize the renormalization group evolution of this relation,
especially to next-to-leading order, requires some effort 
since this study involves a combined study of the (longitudinally) 
polarized and unpolarized evolutions. In this work we present 
a complete NLO study of the evolution equations starting directly from the 
helicity basis. Helicities are in fact the basic parton distributions 
from which other distributions can be built. 

Compared to other implementations, 
in our work we perform a NLO test of the positivity of the
helicity distributions using an ansatz due to Rossi \cite{rossi} which reduces the
evolution equations to an infinite set of recursion relations for some scale invariant coefficients. We have developed a complete implementation of this
algorithm which will be made
available and documented in related work of ours \cite{CaCor}.

Various arguments to validate eq. (\ref{helicities}) 
have been presented in the literature. From our perspective, 
an interesting one has been formulated by Teryaev 
and Collaborators who have tried to establish a link, to leading order, 
between evolution equations and their probabilistic interpretation in 
order to prove Soffer's inequality. Similar arguments hold also in the 
analysis of eq. (\ref{helicities}).

We should remark that a complete probabilistic 
picture exists only for the leading order unpolarized evolution 
\cite{CollinsQiu} and the arguments of \cite{teryaev} are inspired 
by the fact that the subtraction terms (the $x=1$ contributions 
in the expressions of the kernels, where $x$ is Bjorken's variable), 
being positive, once they are combined with the bulk ($x<1$) 
contributions give a form of the evolution equations which are 
diagonal in parton type and resemble ``kinetic'' equations. 
Our arguments, on this issue, are just a 
refinement of this previous and influential analysis. 
 
In the recent literature there has been some attention to this feature of the 
DGLAP evolution, limited to the non-singlet sector, in connection 
with kinetic theory and the "dynamical renormalization group'', in the words of ref.~\cite{devega}. 

All the arguments, so far, go back to some important older work of Collins 
and Qiu who provided an interesting derivation of the (unpolarized) 
DGLAP equation using 
Mueller's formalism of cut diagrams. In their paper \cite{CollinsQiu} the authors reinterpreted 
the DGLAP equation as a kinetic probabilistic equation of Boltzmann type. 
The authors gave no detail on some of the issues concerning the regularization of their 
diagrammatic expansion, on which we will elaborate since we need it for our 
accurate numerical analysis. In our work the Collins-Qiu form of the DGLAP 
equation is interpreted simply as a {\em master equation} rather than a Boltzmann 
equation, given the absence of a 2-to-2 scattering cross section in the 
probabilistic partonic interpretation. A master equation is governed by transition 
probabilities and various formal approximations find their way once this conceptual
step is made. We illustrate, in the spirit of a stochastic approach to the DGLAP dynamics,
how to extract standard differential equations of Kramers-Moyal type for the simplest non-singlet evolutions, those involving valence distributions 
and transverse spin distributions. Our analysis on this point is self-contained 
but purposely short, since a more detailed numerical and formal study 
of this developement is under way \cite{CaCor}. 

We show that the DGLAP dynamics can be described, 
at least in a formal way, by a differential equation of arbitrarily high order. 
Truncations of this expansion to the first few orders provide the usual 
link with the Fokker-Planck approximation, the Langevin equation 
and its path integral version \footnote{For an example of this interplay between differential and 
stochastic descriptions we refer the reader to \cite{BCM}}. 
The picture one should have in mind, at least in this approximation, 
is that of a stochastic (brownian) dynamics of Bjorken's variable $x$ in a 
fictititious time $\log(Q)$, describing the evolution under 
the renormalization group (RG). In this interpretation 
the probability function is the parton distribution itself.

\section{Master Equations and Positivity}
Let's start considering  a generic 1-D master equation for transition
probabilities $w(x|x')$ which we interpret as the probability of making
a transition to a point $x$ given a starting point $x'$ for a given physical system.
The picture we have
in mind is that of a gas of particles making collisions in 1-D and entering the
interval $(x,x + dx)$ with a probability $w(x|x')$ per single transition,
or leaving it with a transition probability $w(x'|x)$.
In general one writes down a master equation
\beq
\frac{\partial }{\partial \tau}f(x,\tau)=\int dx'\left(
w(x|x') f(x',\tau) -w(x'|x) f(x,\tau)\right) dx'.
\eeq
describing the time $\tau$ evolution of the density of the gas
undergoing collisions or the motion of a many replicas of walkers of density
$f(x,\tau)$ jumping with a pre-assigned probability,
according to taste.

The result of Collins a Qiu, who were after a
derivation of the DGLAP equation that could include automatically also the ``edge point''
contributions (or x=1 terms of the DGLAP kernels) is in pointing out the existence of a
probabilistic picture of the DGLAP dynamics.
These edge point terms had been always introduced in the past only by hand and serve to enforce the baryon number sum rule and the momentum sum rule as $Q$, the momentum scale, varies.

The kinetic interpretation was used in \cite{teryaev} to provide an alternative proof
of Soffer's inequality.
We recall that this inequality
\beq
|h_1(x)| < q^+(x)
\eeq
famous by now, sets a bound on the transverse spin distribution $h_1(x)$ in terms of the
components of the positive helicity component of the quarks, for a given flavour.
The inequality has to be respected by the evolution.
We recall that $h_1$, also denoted by the symbol 
\begin{equation}
\Delta _{T}q(x,Q^{2})\equiv q^{\uparrow }(x,Q^{2})-q^{\downarrow }(x,Q^{2}),
\end{equation} 
has the property
of being purely non-singlet and of appearing at leading twist. It is
identifiable in transversely polarized 
hadron-hadron collisions and not in Deep Inelastic Scattering (DIS), where can
appear only through an insertion of the electron mass in the unitarity 
graph of DIS. 

The connection between the Collins-Qiu form of the DGLAP equation and the master equation
is established as follows.
The DGLAP equation, in its original formulation is generically written as

\beq
\frac{d q(x,Q^2)}{d \log( Q^2)} = \int_x^1 \frac{dy}{y} P(x/y)q(y,Q^2),
\eeq
where we are assuming a scalar form of the equation, such as in the non-singlet sector. The generalization to the singlet sector of the arguments given below is, of course, quite straightforward.
To arrive at a probabilistic picture of the equation we start reinterpreting
$\tau=\log (Q^2)$ as a time variable, while the parton density $q(x,\tau)$
lives in a one dimensional (Bjorken) $x$ space.

We recall that the kernels are defined as ``plus'' distributions.
Conservation of baryon number, for instance, is enforced by the addition of
edge-point contributions proportional to $\delta(1-x)$.

We start with the following form of the kernel
\beq
P(z) = \hat{P}(z) - \delta(1-z) \int_0^1 \hat{P}(z)\, dz,
\label{form}
 \eeq
where we have separated  the edge point contributions from the rest
of the kernel, here called $\hat{P}(z)$. This manipulation is understood in all the
equations that follow.
The equation is rewritten in the following form

\beq
\frac{d}{d \tau}q(x,\tau) = \int_x^1 dy \hat{P}\left(\frac{x}{y}\right)\frac{q(y,\tau)}{y}
-\int_0^x \frac{dy}{y}\hat{ P}\left(\frac{y}{x}\right)\frac{q(x,\tau)}{x}
\label{bolz}
\eeq

\begin{figure}[t]
{\centering \resizebox*{6cm}{!}{\rotatebox{0}{\includegraphics{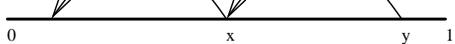}}} \par}
\caption{ The constrained random walk of the parton densities}
\label{walk}
\end{figure}

Now, if we define
\beq
w(x|y)= \frac{\alpha_s}{2 \pi} \hat{P}(x/y)\frac{\theta(y > x)}{y}
\eeq
(\ref{bolz})
becomes a master equation for the probability function $q(x,\tau)$
\beq
\frac{\partial }{\partial \tau}q(x,\tau)=\int dx'\left(
w(x|x') q(x',\tau) -w(x'|x) q(x,\tau)\right) dx'.
\label{masterf}
\eeq
There are some interesting features of this special master equation. Differently from
other master equations, where transitions are allowed from  a given x both toward $y>x$
and $y< x$, in this case, transitions toward x take place only from values $y>x$ and
leave the momentum cell $(x, x+ dx)$ only toward smaller y values (see Fig.(\ref{walk}).

Clearly, this sets a direction of the kinetic evolution of the densities from large
x values toward smaller-x values as $\tau$, the fictitious ``time'' variable, increases.

Probably this is the simplest illustration of the fact that parton densities, at large final evolution scales, are dominated by their
small-x behaviour. As the ``randomly moving partons'' reach the $x\approx 0$ region
of momentum space, they find no space where to go, while other partons tend to pile up
toward the same region from above. This is the picture of a 
random walk biased to move downward (toward the small-x region) and is illustrated in Fig.~(\ref{walk}). 

\section{Probabilistic Kernels}

 We briefly discuss some salient features of the structure of the kernels in this approach and comment on the type 
of regularization involved in order to define them appropriately.  

We recall that unpolarized and polarized kernels, in leading 
order, are given by
\beqa
P^{(0)}_{NS} &=&P_{qq}^{(0)}=C_F\left( \frac{2}{(1-x)_+} -1 - x + 
\frac{3}{2} \delta(1-x)\right) \nonumber\\
P_{qg}^{(0)}&=& 2 T_f\left(x^2 + (1-x)^2)\right) \nonumber \\
P_{gq}^{(0)}&=& C_F \frac{1 + (1 -x)^2}{x}\nonumber \\
P_{gg}^{(0)}&=& 2 N_c\left(\frac{1}{(1-x)_+} +\frac{1}{x} -2 + x(1-x)\right)
 +\frac{\beta_0}{2}\delta(1-x)
\label{stand1}
\eeqa
where
\beq
C_F=\frac{N_C^2 -1}{2 N_C}, \,T_f=T_R n_f= \frac{1}{2} n_f,
\,\,\beta_0=\frac{11}{3} N_C  - \frac{4}{3} T_f \nonumber\\
\eeq
and
\beqa
\Delta P^{(0)}_{NS} &=&\Delta P_{qq}^{(0)}\nonumber\\
\Delta P_{qq}^{(0)}&=&C_F\left( \frac{2}{(1-x)_+} - 1 - x +
 \frac{3}{2}\delta(1-x)\right)\nonumber \\
\Delta P_{qg}^{(0)}&=& 2 T_f( 2 x - 1) \nonumber \\
\Delta P_{gq}^{(0)}&=& C_F (2 -x)\nonumber \\
\Delta P_{gg}^{(0)}&=& 2 N_c\left(\frac{1}{(1-x)_+} - 2 x + 1\right)
+\delta(1-x)\frac{\beta_0}{2},
\label{stand2}
\eeqa

while the LO transverse kernels are given by
\beq
\Delta_T P^{(0)}_{qq}=C_F\left( \frac{2 }{(1-x)_+} - 2  
+ \frac{3}{2}\delta(1-x)\right).
\eeq

The unpolarized kernels should be compared with the Collins-Qiu form 
\beqa
P_{qq}&=&\gamma_{qq} -\delta(1-x) \int_0^1 dz \gamma_{qq}
\nonumber \\
P_{gg}&=&\gamma_{gg} - 
\left(n_f \int_0^1 {dz} \gamma_{qg} +\frac{1}{2}\int_0^1 dz 
\gamma_{gg}\right) \delta(1-x) \nonumber \\
P_{qg}&=& \gamma_{qg}\nonumber \\
P_{gq}&=& \gamma_{gq}\nonumber \\
\label{cq1}
\eeqa
where
\beqa
\gamma_{qq}&=& C_F \left(\frac{2}{1-x} -1 - x 
\right)\nonumber \\
\gamma_{qg} &=& (2 x -1)\nonumber \\
\gamma_{gq} &=& C_F(2 - x) \nonumber \\
\gamma_{gg} &=& 2 N_c\left( \frac{1}{1-x} +\frac{1}{x} -2 + x(1-x) \right). \nonumber \\
\label{cq2}
\eeqa
These kernels need a suitable regularization to be well defined. 
Below we will analize the implicit regularization 
underlying eq.~(\ref{cq1}). One observation is however almost immediate: 
the component $P_{gg}$ is not of the form given by eq.~(\ref{form}). 
In general, therefore, in the singlet case, the generalization of 
eq.~(\ref{form}) is given by 
\beq
P(x)=\hat{P}_1(x) - \delta(1-x)\int_0^1 \hat{P}_2(z) dz
\eeq
and a probabilistic interpretation is more complex compared to the non-singlet 
case and has been discussed in the original literature \cite{CollinsQiu}. 

\section{Convolutions and Master Form of the Singlet}
Distributions are folded with the kernels and the result rearranged
in order to simplify the structure of the equations. Since
in the previous literature this is done in a rather involuted way 
\cite{Hinchliffe} 
we provide here a simplificaton, from which the equivalence of the 
various forms of the kernel, in the various regularizations adopted, will be 
apparent.     
All we need is the simple relation
\beq
\int_x^1 \frac{dy}{y (1-y)_+}f(x/y)=\int_x^1\frac{dy}{y}
\frac{ yf(y) - x f(x)}{y-x} +\log(1-x) f(x)
\label{simplerel}
\eeq
in which, on the right hand side, regularity of both the first 
and the second term is explicit. For instance, the evolution equations become
\beqa
\frac{d q}{d \log(Q^2)} &=& 2 C_F 
\int\frac{dy}{y}\frac{ y q(y) - x q(x)}{y-x} +2 C_F \log(1-x)\, q(x) -
\int_x^1\frac{dy}{y}\left( 1 + z\right)q(y) + 
\frac{3}{2} C_F q(x) \nonumber \\
&& + n_f\int_x^1\frac{dy}{y}
\left( z^2 +(1-z)^2\right)g(y)\nonumber \\
\frac{d g}{d \log(Q^2)} &=& 
C_F \int_x^1\frac{dy}{y}\frac{1 +(1-z)^2}{z}q(y)
+ 2 N_c \int_x^1\frac{dy}{y}
\frac{ y f(y) - x f(x)}{y-x}g(y) 
\nonumber \\
&& + 2 N_c \log(1-x) g(x) 
+2 N_c\int_x^1 \frac{dy}{y}\left( \frac{1}{z} -2 + z(1-z)\right)g(y) + 
\frac{\beta_0}{2}g(x) \nonumber \\
\label{standard}
\eeqa
with $z\equiv x/y$. The same simplified form is obtained 
from the probabilistic version, having defined a suitable regularization 
of the edge point singularities in the integrals over the components 
$\gamma_{f f'}$ in eq. (\ref{cq2}). The canonical expressions of the 
kernels (\ref{stand2}), expressed in terms of ``+'' distributions, can also be rearranged to look like their equivalent probabilistic 
form by isolating the edge-point contributions hidden in their ``+''
distributions. We get the expressions

\beqa
{P_{qq}^{(0)}}_{NS} &=&P_{qq}^{(0)}=C_F\left( \frac{2}{(1-x)} -1 - x\right) - 
\left(C_F\int_0^1 \frac{dz}{1-z} -\frac{3}{2}\right) \delta(1-x) \nonumber\\
P_{gg}^{(0)}&=& 2 N_c\left(\frac{1}{(1-x)} +\frac{1}{x} -2 + x(1-x)\right)
-\left(2 N_c \int_0^1 \frac{dz}{1-z}-\frac{\beta_0}{2}\right)\delta(1-x)
\label{stand1}
\eeqa

and
\beqa
\Delta P_{qq}^{(0)}&=&C_F\left( \frac{2}{(1-x)} - 1 - x\right) -
C_F\left( \int_0^1 \frac{dz}{1-z} - \frac{3}{2}\right)\delta(1-x)\nonumber \\
\Delta P_{gg}^{(0)}&=& 2 N_c\left(\frac{1}{1-x} - 2 x + 1\right)
-\left(2 N_c  \int_0^1 \frac{dz}{1-z}  -\frac{\beta_0}{2}\right) \delta(1-x),
\label{stand2}
\eeqa
the other expressions remaining invariant. In appendix A we provide some 
technical details on the equivalence between the convolutions obtained 
using these kernels with the standard ones.

A master form of the singlet (unpolarized) equation 
is obtained by a straightforward change of variable in the decreasing 
terms. We obtain 
\beqa
\frac{d q}{d \tau}&=&\int_x^{1 - \Lambda}\frac{dy}{y}\gamma_{qq}(x/y)q(y) 
-\int_0^{x - \Lambda}\frac{dy}{y}\gamma_{qq}(y/x) q(x)\nonumber \\
\frac{d g}{d \tau}&=&\int_x^{1 - \Lambda}\frac{dy}{y}\gamma_{gg}(x/y)
- n_f\int_0^x\gamma_{qg}(y/x) g(x)  \nonumber \\
&& -\frac{1}{2} \int_\Lambda^{x - \Lambda}\gamma_{gg}(y/x) g(x) + 
\int_x^1\frac{dy}{y}\gamma_{gq}(x/y)q(y)
\eeqa 
with a suitable (unique) cutoff $\Lambda$ needed to cast the 
equation in the form (\ref{standard}). A discussion of this aspect is 
left in appendix B.  
The (regulated) transition probabilities are then given by 
\beqa
w_{qq}(x|y)&=&\gamma_{qq}(x/y)\frac{\theta(y>x)\theta(y< 1 -\Lambda)}{y}\nonumber \\  
w_{qq}(y|x)&=&\gamma_{qq}(y/x)\frac{\theta(y< x- \Lambda) \theta(y>0)}{x} \nonumber \\ 
w_{gg}(x|y) &=&\gamma_{gg}(x/y)\frac{\theta(y>x)\theta(y< 1 -\Lambda)}{y} \nonumber \\ 
w_{qq}(y|x)&=&\left(n_f\gamma_{qg}(y/x)
-\frac{1}{2}\gamma_{gg}(y/x)\right)\frac{\theta(y< x- \Lambda) \theta(y>0)}{x} \nonumber \\
w_{gq}(y|x)&=&\gamma_{gq}(x/y)\frac{\theta(y>x)\theta(y< 1 -\Lambda)}{y}\nonumber \\  
w_{gq}(x|y)&=& 0, \nonumber \\
\eeqa
as one can easily deduct from the form of eq. (\ref{masterf}).

\section{A Kramers-Moyal Expansion for the DGLAP Equation}
Kramers-Moyal (KM) expansions of the master equations (backward or forward)
are sometimes useful in order to gain insight into the master equation itself,
since they may provide a complementary view of the underlying dynamics.

The expansion allows to get rid of the integral which characterizes the master
equation, at the cost of introducing a differential operator of arbitrary
order. For the approximation to be useful, one has to stop the expansion after
the first few orders. In many cases this turns out to be possibile.
Examples of processes of this type are special Langevin processes and
processes described by a Fokker-Planck operator. In these cases
the probabilistic interpretation allows us to write down a fictituous
lagrangean, a corresponding path integral and solve for the
propagators using the Feynman-Kac fomula. 
 For definitess we take the integral to cover all
the real axis in the variable $x'$ 

\beq
\frac{\partial }{\partial \tau}q(x,\tau)=\int_{-\infty}^{\infty} dx'\left(
w(x|x') q(x',\tau) -w(x'|x) q(x,\tau)\right) dx'.
\eeq

As we will see below, in the DGLAP case some
modifications to the usual form of the KM expansion will appear. 
At this point we perform a KM expansion of the equation in the usual way.
We make the substitutions in the master equation $y\to x-y$ in the first term and
$y\to x + y$ in the second term

\beq
\frac{\partial }{\partial \tau}q(x,\tau)=\int_{-\infty}^{\infty} dy\left(
w(x|x -y) q(x - y,\tau) -w(x + y|x) q(x,\tau)\right) ,
\eeq
identically equal to
\beq
\frac{\partial }{\partial \tau}q(x,\tau)=\int_{-\infty}^{\infty} dy\left(
w(x + y - y'|x -y') q(x - y',\tau) - w(x + y'|x) q(x,\tau)\right),
\label{shift}
\eeq
with $y=y'$. First and second term in the equation above differ by a shift (in $-y'$) and can be related using a Taylor (or KM) expansion of the first term

\beqa
\frac{\partial }{\partial \tau}q(x,\tau)&=&\int_{-\infty}^{\infty} dy \sum_{n=1}^\infty
\frac{(-y)^n}{n!}\frac{\partial^n}{\partial x^n}\left( w(x + y|x)
q(x,\tau)\right)
\eeqa
where the $n=0$ term has canceled between the first and the second
contribution coming from (\ref{shift}). The result can be
written in the form
\beq
\frac{\partial }{\partial \tau}q(x,\tau)= \sum_{n=1}^\infty
\frac{(-y)^n}{n!}\frac{\partial^n}{\partial x^n}\left( a_n(x)q(x,\tau)\right)
\eeq
where
\beq
a_n(x)=\int_{-\infty}^{\infty} dy(y-x)^n w(y|x).
\eeq

In the DGLAP case we need to amend the former derivation, due to the
presence of boundaries $( 0 < x < 1)$ in the Bjorken variable $x$. For simplicity 
we will focus on the non-singlet case. 
We rewrite the master equation using the same change of variables used above 

\beqa
\frac{\partial}{\partial\tau}q(x,\tau) &=& \int_x^1 dy w(x|y)q(y,\tau) - 
\int_0^x dy w(y|x) q(x,\tau)
\nonumber \\
&& -\int_0^{\alpha(x)} dy w(x+y|x)* q(x,\tau)+ 
\int_0^{-x} dy w(x+y|x)q(x,\tau),
\eeqa

where we have introduced the simplest form of the Moyal product 
\footnote{A note for noncommutative geometers: 
this simplified form is obtained for a dissipative dynamics when the 
{\bf p}'s of phase space are replaced by constants. Here we have only one variable: $x$} 
 
\beq
w(x+y|x)*q(x)\equiv w(x+y|x) e^{-y \left(\overleftarrow{\partial}_x + 
\overrightarrow{\partial}_x\right)} q(x,\tau)
\eeq
and $\alpha(x) =x-1$. 
The expansion is of the form 
\beqa
\frac{\partial}{\partial \tau}q(x,\tau)=\int_{\alpha(x)}^{-x}dy\,  w(x+y|x)q(x,\tau) - 
\sum_{n=1}^{\infty}\int_0^{\alpha(x)}dy \frac{(-y)^n}{n!}{\partial_x}^n
\left(w(x+y|x)q(x,\tau)\right)
\label{expans}
\eeqa
which can be reduced to a differential equation of arbitrary order using simple manipulations. 
We recall that the Fokker-Planck approximation is obtained stopping the expansion at 
the second order 
\beq
\frac{\partial}{\partial \tau}q(x,\tau)= a_0(x) -\partial_x\left(a_1(x) q(x)\right) + 
\frac{1}{2}\partial_x^2\left( a_2(x) q(x,\tau)\right)
\label{fpe}
\eeq
with 
\beq
a_n(x)=\int dy \,y^n\, w(x+y,x)
\eeq
being moments of the transition probability function $w$.
Given the boundary conditions on the Bjorken variable x, even in the 
Fokker-Planck approximation, the Fokker-Planck version of the DGLAP equation 
is slightly more involved than Eq. (\ref{fpe}) and the coefficients $a_n(x)$ need to be redefined.

\section{The Fokker-Planck Approximation}
The probabilistic interpretation of the DGLAP equation motivates us to investigate the role of the Fokker-Planck (FP) approximation to the equation and its possible practical use. 
We should start by saying a word of caution regarding this expansion. 

In the context of a random walk, an all-order derivative expansion of the master equation 
can be arrested to the first few terms either if the conditions of Pawula's 
theorem are satisfied -in which case the FP approximation turns out to be exact- 
or if the transition probabilities show an exponential decay above 
a certain distance allowed to the random walk. Since the DGLAP kernels 
show only an algebraic decay in x, 
and there isn't any explicit scale in the kernel themselves, 
the expansion is questionable. However, from a formal viewpoint, it is still allowed. 
With these caveats in mind we proceed to investigate the features of this expansion. 
  
We redefine 

\beqa
\tilde{a}_0(x) &=& \int_{\alpha(x)}^{-x} dy w(x+y|x)q(x,\tau) \nonumber \\
a_n(x) &=&\int_0^{\alpha(x)} dy y^n w(x+y|x) q(x,\tau) \nonumber \\
\tilde{a}_n(x)&=&\int_0^{\alpha(x)}
dy y^n {\partial_x}^n \left(w(x+y|x)q(x,\tau)\right) \,\,\,n=1,2,...
\eeqa
For the first two terms $(n=1,2)$ one can easily work out the relations
\beqa
\tilde{a}_1(x) &=&\partial_x a_1(x) - \alpha(x) \partial_x \alpha(x)
w(x + \alpha(x)|x)q(x,\tau) \nonumber \\
\tilde{a}_2(x) &=&\partial_x^2 a_2(x) - 
2 \alpha(x) (\partial_x \alpha(x))^2 w(x+ \alpha(x)|x) q(x,\tau) - 
\alpha(x)^2 \partial_x\alpha(x)
\partial_x\left( w(x+ \alpha(x)|x)q(x,\tau)\right)\nonumber \\ 
&& - \alpha^2(x)\partial_x \alpha(x) \partial_x\left( w(x + y|x)q(x,\tau)\right)|_{y=\alpha(x)}
\eeqa

Let's see what happens when we arrest the expansion (\ref{expans}) to the first 3 terms.
The Fokker-Planck version of the equation is obtained by including in the approximation
only $\tilde{a}_n$ with $n=0,1,2$.

The Fokker-Planck limit of the (non-singlet) equation is then given by
\beq
\frac{\partial}{\partial \tau}q(x,\tau) = \tilde{a}_0(x)
+\tilde{a}_1(x) - \frac{1}{2} \tilde{a}_2(x)
\eeq
which we rewrite explicitely as 
\beqa
\frac{\partial}{\partial \tau}q(x,\tau) &=& 
C_F\left(\frac{85}{12} +\frac{3}{4 x^4} - \frac{13}{3 x^3} +
\frac{10}{x^2} -\frac{12}{x}
+ 2\log\left(\frac{1-x}{x}\right)\right)q(x)\nonumber \\
&& +C_F\left( 9 - \frac{1}{2 x^3} +\frac{3}{x^2} -\frac{7}{x} 
-\frac{9}{2}\right) \partial_x q(x,\tau) \nonumber \\
&& + C_F\left(\frac{9}{4} +\frac{1}{8 x^2} -\frac{5}{6 x} -
\frac{5 x}{2}  +\frac{23 x^2}{24}\right) \partial_x^2 q(x,\tau).
\eeqa

A similar approach can be followed also for other cases, for which a 
probabilistic picture (a derivation of Collins-Qiu type) has not been established yet, 
such as for $h_1$. 
We describe briefly how to proceed in this case. 

First of all, we rewrite the evolution equation for the transversity 
in a suitable master form. This is possible since the subtraction terms 
can be written as integrals of a positive function. A possibility is 
to choose the transition probabilities
\beqa
w_1[x|y] &=& \frac{C_F}{y}\left(\frac{2}{1- x/y} - 2 \right)
\theta(y>x) \theta(y<1)\nonumber \\
w_2[y|x] &=& \frac{C_F}{x} \left(\frac{2}{1- y/x} - \frac{3}{2}\right)
\theta(y > -x)\theta(y<0)
\nonumber \\
\eeqa
which reproduce the evolution equation for $h_1$ in master form

\beq
\frac{d h_1}{d \tau}= \int_0^1 dy w_1(x|y)h_1(y,\tau) 
-\int_0^1 dy w_2(y|x) h_1(x,\tau).
\eeq
The Kramers-Moyal expansion is derived as before, with some slight
modifications. The result is obtained introducing an intermediate cutoff which is 
removed at the end. In this case we get 
\beqa
\frac{d h_1}{d\tau} &=& C_F\left( \frac{17}{3}
-\frac{2}{3 x^3} + \frac{3}{x^2} - \frac{6}{x}
+ 2 \log\left(\frac{1-x}{x}\right)\right) h_1(x,\tau) \nonumber \\
&& + C_F\left( 6 + \frac{2}{3 x^2} -\frac{3}{x} 
- \frac{11 x }{3}\right)\partial_x h_1(x,\tau) \nonumber \\
&& + C_F\left( \frac{3}{2} - \frac{1}{3 x } 
- 2 x + \frac{5 x^2}{6}\right)\partial_x^2 h_1(x,\tau). \nonumber 
\eeqa

Notice that compared to the standard Fokker-Planck approximation, the boundary now 
generates a term on the left-hand-side of the equation proportional to $q(x)$ 
which is absent in eq. (\ref{fpe}). This and higher order approximations to the 
DGLAP equation can be studied systematically both analytically and numerically 
and it is possible to assess the validity of the approximation \cite{CaCor}.

\section{Helicities to LO}
As we have mentioned above, an interesting version of the usual DGLAP equation involves
the helicity distributions.

We start introducing \cite{teryaev} the DGLAP kernels for fixed
helicites
$P_{++}(z)=(P(z)+\Delta P(z))/2$ and $ P_{+-}(z)=(P(z)-\Delta P(z))/2$
which will be used below. $P(z)$
denotes (generically) the unpolarized kernels, while the $\Delta P(z)$ are the longitudinally polarized ones. These definitions, throughout the paper, are meant to be
expanded up to NLO, the order at which our numerical analysis holds.

\begin{figure}
{\centering \resizebox*{12cm}{!}{\rotatebox{-90}{\includegraphics{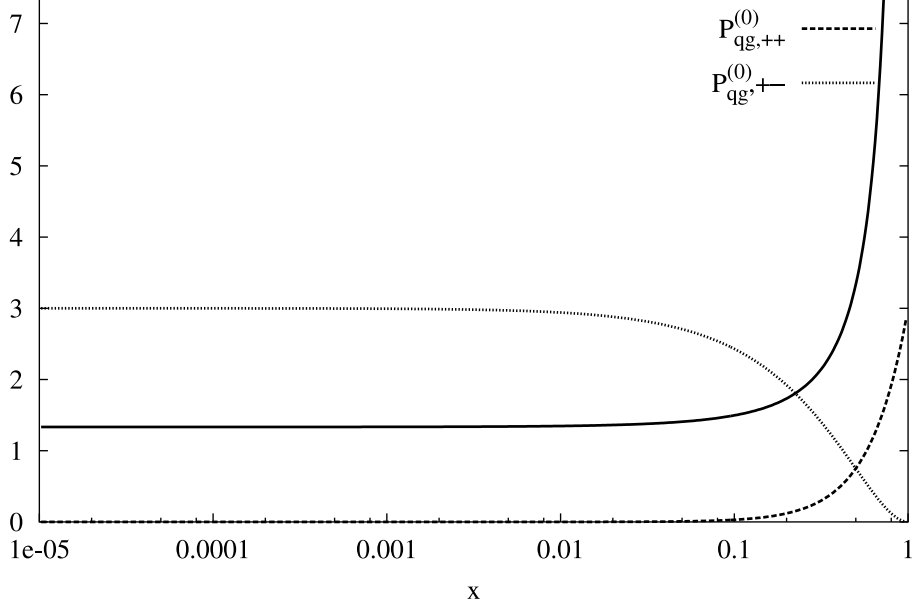}}} \par}
\caption{LO kernels (\protect\( qq\protect \) and \protect\( qg\protect \))
in the helicity basis.}
\label{p0qx}
\end{figure}

 The equations, in the helicity basis, are 

\begin{eqnarray}
{dq_+(x) \over{dt}}=
{\alpha_s \over {2 \pi}} (P_{++}^{qq} ({x \over y}) \otimes q_+(y)+
P_{+-}^{qq} ({x \over y}) \otimes q_-(y)  \nonumber \\
+P_{++}^{qg} ({x \over y}) \otimes g_+(y)+
P_{+-}^{qg} ({x \over y}) \otimes g_-(y)),
\nonumber \\
{dq_-(x) \over{dt}}=
{\alpha_s \over {2 \pi}} (P_{+-} ({x \over y}) \otimes q_+(y)+
P_{++} ({x \over y}) \otimes q_-(y) \nonumber \\
+P_{+-}^{qg} ({x \over y}) \otimes g_+(y)+
P_{++}^{qg} ({x \over y}) \otimes g_-(y)),  \nonumber \\
{dg_+(x) \over{dt}}=
{\alpha_s \over {2 \pi}} (P_{++}^{gq} ({x \over y}) \otimes q_+(y)+
P_{+-}^{gq} ({x \over y}) \otimes q_-(y) \nonumber \\
+P_{++}^{gg} ({x \over y}) \otimes g_+(y)+
P_{+-}^{gg} ({x \over y}) \otimes g_-(y)),  \nonumber \\
{dg_-(x) \over{dt}}=
{\alpha_s \over {2 \pi}} (P_{+-}^{gq} ({x \over y}) \otimes q_+(y)+
P_{++}^{gq} ({x \over y}) \otimes q_-(y) \nonumber \\
+P_{+-}^{gg} ({x \over y}) \otimes g_+(y)+
P_{++}^{gg} ({x \over y}) \otimes g_-(y)).
\label{hs}\end{eqnarray}
The non-singlet (valence) analogue of this equation is also easy to
write down
\begin{eqnarray}
{dq_{+, V}(x) \over{dt}}=
{\alpha_s \over {2 \pi}} (P_{++} ({x \over y}) \otimes q_{+,V}(y)+
P_{+-} ({x \over y}) \otimes q_{-,V}(y)), \nonumber \\
{dq_{-,V}(x) \over{dt}}=
{\alpha_s \over {2 \pi}} (P_{+-} ({x \over y}) \otimes q_{+,V}(y)+
P_{++} ({x \over y}) \otimes q_{-,V}(y)).
\label{h}\end{eqnarray}
where the $q_{\pm,V}=q_\pm - \bar{q}_\pm$ are the valence components of fixed helicites.
The kernels in this basis are given by 
\beqa
P_{NS\pm,++}^{(0)} &=&P_{qq, ++}^{(0)}=P_{qq}^{(0)}\nonumber \\
P_{qq,+-}^{(0)}&=&P_{qq,-+}^{(0)}= 0\nonumber \\
P_{qg,++}^{(0)}&=& n_f x^2\nonumber \\
P_{qg,+-}&=& P_{qg,-+}= n_f(x-1)^2 \nonumber \\
P_{gq,++}&=& P_{gq,--}=C_F\frac{1}{x}\nonumber \\ 
P_{gg,++}^{(0)}&=&P_{gg,++}^{(0)}= N_c
\left(\frac{2}{(1-x)_+} +\frac{1}{x} -1 -x - x^2 \right) +{\beta_0}\delta(1-x) \nonumber \\
P_{gg,+-}^{(0)}&=& N_c
\left( 3 x +\frac{1}{x} -3 - x^2 \right) 
\label{stand1}
\eeqa

Taking linear combinations of these equations (adding and subtracting),
one recovers the usual evolutions for unpolarized $q(x)$ and longitudinally
polarized $\Delta q(x)$ distributions.
We recall that the unpolarized distributions, the polarized and the transversely
polarized $q_T(x)$ are related by

\begin{figure}
{\centering \resizebox*{12cm}{!}{\rotatebox{-90}{\includegraphics{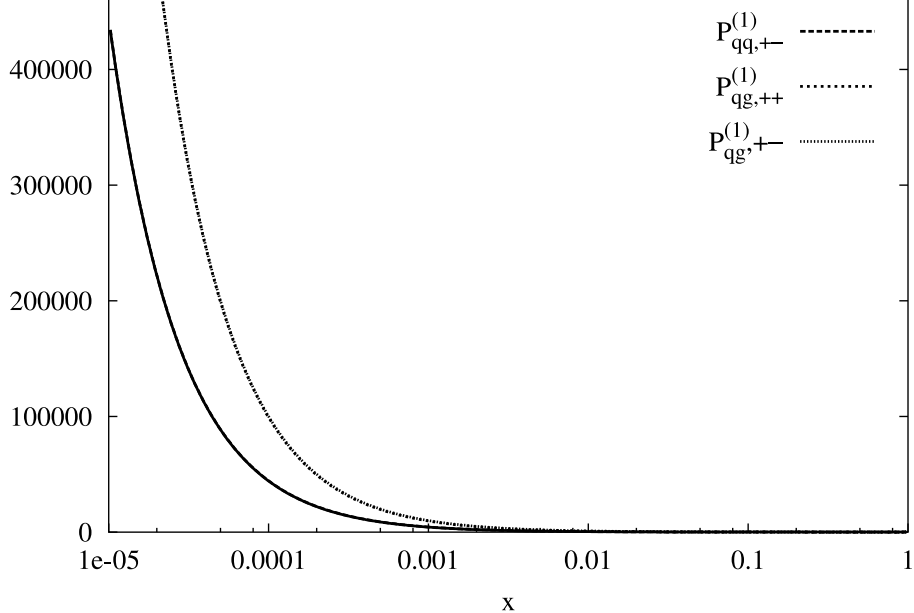}}} \par}

\caption{NLO kernels (\protect\( qq\protect \) and \protect\( qg\protect \))
in the helicity basis.}
\label{p1qx}
\end{figure}

\beqa
q(x)&=&q_+(x) + q_-(x)=q_{+T}(x) + q_{-T}(x)\nonumber \\
\Delta q(x)&=& q_+(x) - q_-(x)
\eeqa
at any $Q$ of the evolution and, in particular, at the boundary of the
evolution. 

\begin{figure}
{\centering \resizebox*{12cm}{!}{\rotatebox{-90}{\includegraphics{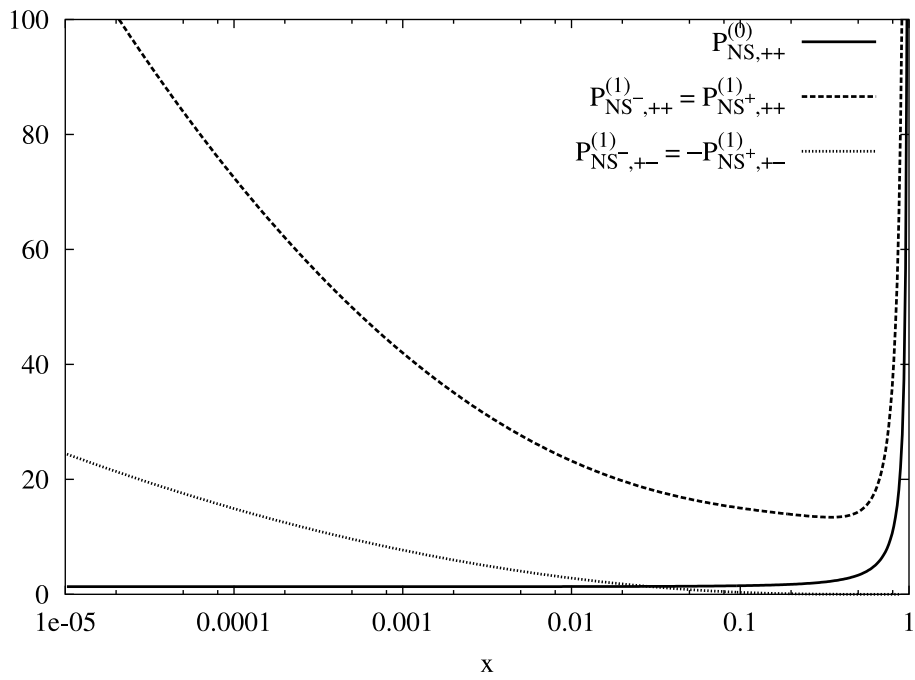}}} \par}
\caption{Nonsinglet kernels in the helicity basis.}
\label{pns}
\end{figure}

Similar definition have been introduced for the gluon sector
with $G_\pm(x)$ denoting the fixed helicities of the gluon distributions with
$\Delta g(x) = g_+(x) - g_-(x)$ and  $g(x)= g_+(x) + g_-(x)$ being the
corresponding longitudinal asymmetry and the unpolarized density respectively.

\begin{figure}
{\centering \resizebox*{12cm}{!}{\rotatebox{-90}{\includegraphics{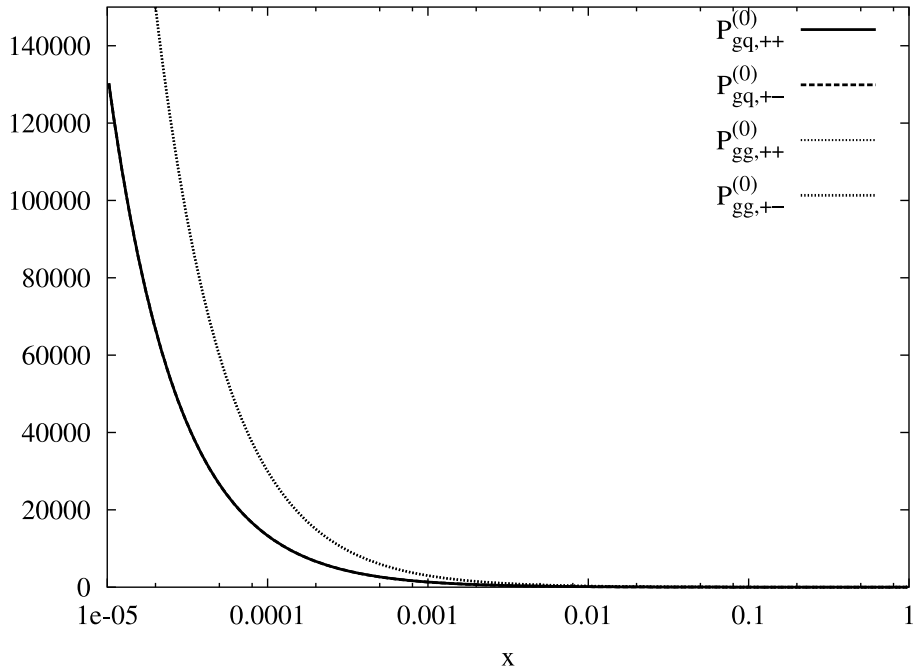}}} \par}

\caption{LO kernels (\protect\( gq\protect \) and \protect\( gg\protect \))
in the helicity basis.}
\label{p0gx}
\end{figure}

\section{Summary of Positivity Arguments}

Let's recapitulate here the basic arguments \cite{teryaev} that are
brought forward in order to prove the positivity of the evolution to NLO.

If
\beq
|\Delta P(z)| \leq P(z), z < 1
\label{condi}
\eeq
then both kernels $P_{++}(z)$ and $P_{+-}(z)$ are positive as far as $z < 1$.

The singular contributions at $z=1$, which appear as subtraction terms
in the evolution and which could, in principle, alter positivity, appear
only in diagonal form, which means that they are only contained
in $P_{++}$, multiplied by the single functions $q_+(x)$ or $q_-(x)$

\beq
{dq_+(x) \over{dt}}=
{\alpha_s \over {2 \pi}} P_{++}^{qq} ({x \over y}) \otimes q_+(y) + . . .
\label{qplus}
\eeq
\beq
{dq_-(x) \over{dt}}=
{\alpha_s \over {2 \pi}} P_{++}^{qq} ({x \over y}) \otimes q_+(y) + . . .
\eeq
\beq
{d g_+(x) \over{dt}}=
{\alpha_s \over {2 \pi}} P_{++}^{gg} ({x \over y}) \otimes g_+(y) + . . .
\eeq

Let's focus just on the equation for $q_+$ (\ref{qplus}). Rewriting the diagonal contribution as a master equation
\beq
{dq_+(x) \over{dt}}=\int dx'\left(
w_{++}(x|x') q_+(x',\tau) -w_{++}(x'|x) q_+(x,\tau)\right) dx' + . . .
\label{formx}
\eeq
in terms of a transition probability
\beq
w_{++}(x|y)= \frac{\alpha_s}{2 \pi} \hat{P}_{++}(x/y)\theta(y > x)
\eeq
which can be easily established to be positive, as we are going to show rigorously below, 
as far as all the remanining
terms (the ellipses) are positive. We have performed a detailed 
numerical analysis to show the positivity of the contributions at $x=1$. 
 
This last condition is also clearly satisfied, since the
$\delta(1-z)$ contributions appear only in $P_{++}$ and are diagonal in
the helicity of the various flavours (q, g). 
For a rigorous proof of the positivity of the solutions of master equations 
we proceed as follows. 

\begin{figure}
{\centering \resizebox*{12cm}{!}{\rotatebox{-90}{\includegraphics{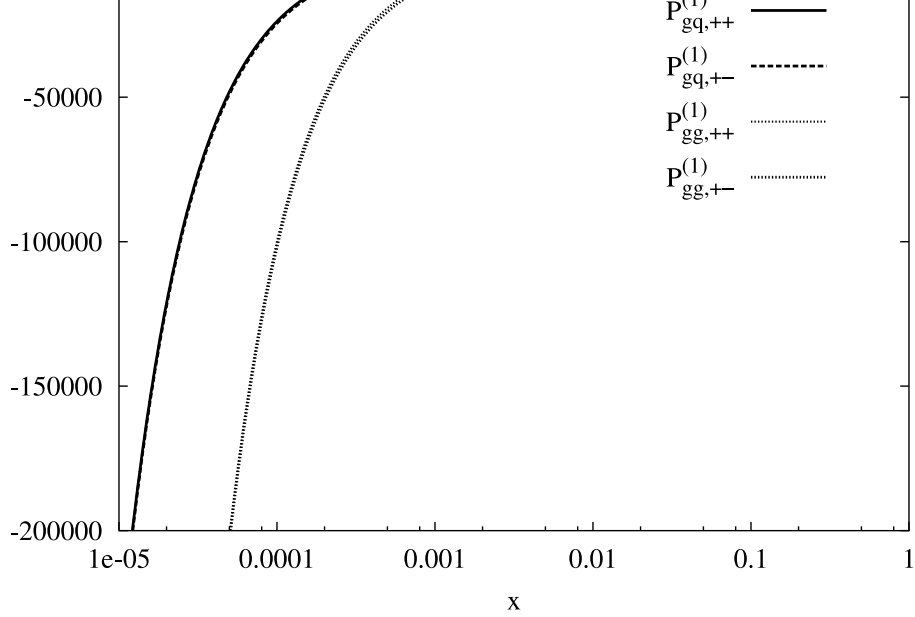}}} \par}

\caption{NLO kernels (\protect\( gq\protect \) and \protect\( gg\protect \))
in the helicity basis.}
\label{p1gx}
\end{figure}

Let $q(x,\tau)$  be a positive distribution for 
$\tau < \tau_c$ and let us assume that it vanishes at $\tau=\tau_c$, after which 
it turns negative. We also assume that the evolution of 
$q(x,\tau)$ is of the form (\ref{form}) with positive 
transition probabilities $w(x|y)$ and $w(y|x)$. Notice that since 
the function is continuous together with its first derivative and decreasing, 
continuity of its first derivative will require $\frac{d q(x,\tau)}{d \tau} < 0$ at 
$\tau=\tau_c$ and in its neighbor. However, eq. (\ref{form}) requires 
that at $\tau=\tau_c$   
\beq
{dq(x) \over{dt}}=\int dx'
w (x|x') q(x',\tau) 
\label{form1}
\eeq

which is positive, and we have a contradiction. 
We can picture the evolution in $\tau$ of these functions as a family 
of curves getting support to smaller and smaller x-values as $\tau$ grows 
and being almost vanishing at intermediate and large x values. 
We should mention that this proof does not require a complete 
probabilistic picture of the evolution, but just the positivity of the 
bulk part of the kernels, the positivity 
of the edge point subtractions and their diagonality in flavour. 
From Figs. (\ref{p0qx})  and (\ref{p0gx}) it is also evident that the leading order 
kernels are positive, together with the $qg$ and $qq$ (Fig. (\ref{p1qx})) sectors. 

The edge point contributions, generating the ``subtraction terms'' in the master 
equations for the ``++'' components of the kernels are positive, 
as is illustrated in 3 Tables included in Appendix A. 
There we have organized these terms in the form $\sim C\delta(1-x)$ with
\beq
C=-\log(1- \Lambda) A + B 
\label{sub}
\eeq
with A and B being numerical coefficients depending on the number 
of flavours included in the kernels. Notice that the subtraction terms are 
always of the form (\ref{sub}), with the (diverging) logarithmic contribution 
($\sim \int_0^\Lambda dz/(1-z)$) regulated by a cutoff. This divergence 
in the convolution cancels when these terms are combined with the divergence at 
$x=1$ of the first term of the master equation for all the relevant components 
containing ``+'' distributions. 
It is crucial, however, to establish positivity of the evolution of the 
helicities that the boundary conditions on the evolution
$|\Delta q(x,Q_0^2)| \leq q(x,Q_0^2)$ be satisfied. 
Initial conditions have this special property, in most of models, and the proof 
of positivity of all the distributions therefore holds at any $Q$. 

\begin{figure}
{\centering \resizebox*{12cm}{!}{\rotatebox{-90}{\includegraphics{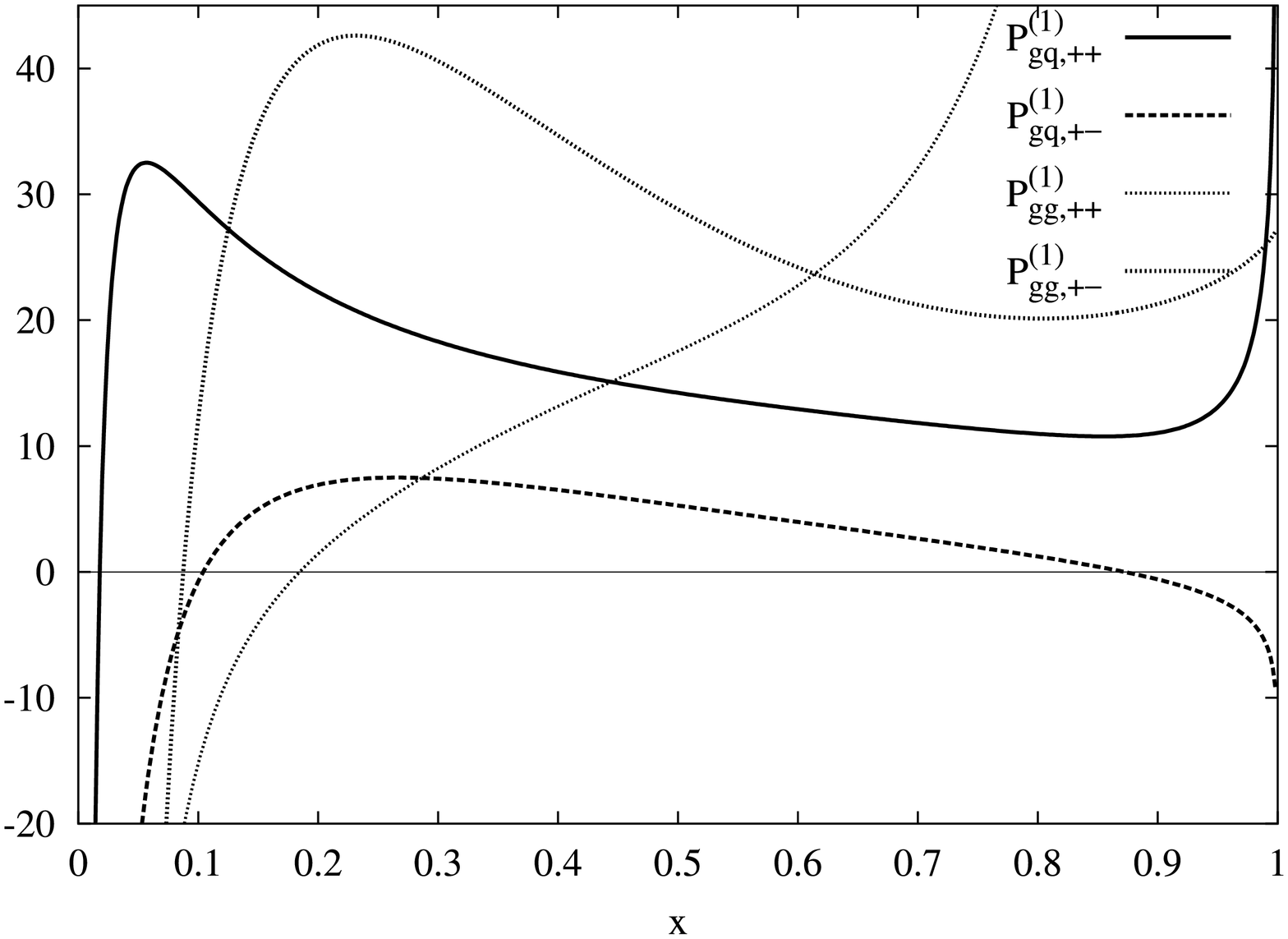}}} \par}
\caption{The crossing of NLO kernels (\protect\( gq\protect \) and \protect\( gg\protect \))
in the helicity basis.}
\label{p1gx_nonlog}
\end{figure}

As we move to NLO, the pattern gets more complicated.  
In fact, from a numerical check, one can see that some NLO kernels 
turn to be negative, including the unpolarized kernels and the 
helicity kernels, while others (Fig. (\ref{pns})) are positive.  
One can also notice 
the presence of a crossing of several helicity components in the $gq$ and $gg$ 
sectors (see Fig. (\ref{p1gx_nonlog})) at larger x values, while in the 
small-x region some components turn negative (Fig. (\ref{p1gx})). 
There is no compelling proof of positivity, in this case, either 
than that coming from a direct numerical analysis.

\section{NLO Numerical Tests using Recursion Relations }
We have seen that 
master forms of evolution equations, for evolutions of all kinds, when found, 
can be used to establish positivity of the evolution itself. 

The requirements  have been spelled out above and can be
summarized in the following points: 1) diagonality of the decreasing terms,
2) initial positivity of the distributions, 3) positivity
of the remaining (non diagonal) kernels. 
As we have also seen, some of these conditions are not satisfied by the NLO evolution.  

In order then to proceed with a numerical test of the inequality 
we have decided to work directly in x-space, using a specific ansatz 
which summarizes the NLO evolution in a rather compact form. 

This ansatz, which we will illustrate below, 
reduces the evolution equations to a suitable
set of recursion relations \cite{rossi}, \cite{storrow}.

In this ansatz, 
the NLO expansion of the distributions in the DGLAP equation is generically given by
\beq
f(x,Q^2)=\sum_{n=0}^{\infty} \frac{A_n(x)}{n!}\log^n
\left(\frac{\alpha(Q^2)}{\alpha(Q_0^2)}\right) +
\alpha(Q^2)\sum_{n=0}^\infty \frac{B_n(x)}{n!}\log^n
\left(\frac{\alpha(Q^2)}{\alpha(Q_0^2)}\right)
\label{expansion}
 \eeq
where we assume a short hand matrix notation for all the convolution products.
Notice that $f(x,Q^2)$ stands for a vector having as components all the helicities
of the various flavours  $(q_\pm,G_\pm)$.
The ansatz implies a tower of recursion relations 
once the running coupling is kept into account \cite{gordon, CCG}

\beq
A_{n+1}(x) =  -\frac{2}{\beta_0}P^{(0)}\otimes A_n(x)
\eeq
to leading order
and
\beqa
B_{n+1}(x) & = & - B_n(x)- \left(\frac{\beta_1}{4 \beta_0} A_{n+1}(x)\right)
- \frac{1}{4 \pi\beta_0}P^{(1)}\otimes A_n(x) -\frac{2}{\beta_0}P^{(0)}
\otimes B_n(x) \nonumber \\
 & = &  - B_n(x) + \left(\frac{\beta_1}{2 \beta_0^2}P^{(0)}\otimes A_n(x)\right)
 \nonumber \\
&&- \frac{1}{4 \pi\beta_0}P^{(1)}\otimes A_n(x) -\frac{2}{\beta_0}P^{(0)}
\otimes B_n(x),  \nonumber \\
\label{recur1}
\eeqa
to NLO, relations which are solved with the initial condition $B_0(x)=0$.
The initial conditions for the coefficients $ A_0(x)$ and $B_0(x)$ are specified 
with $q(x,Q_0^2)$ as a leading order ansatz for the initial
distribution
\beq
A_0(x)= \delta(1-x)\otimes q(x,Q_0^2)\equiv q_0(x)
\eeq
which also requires $B_0(x)=0$, since we have to
satisfy the boundary condition \cite{CCG}
\beq
A_0(x) + \alpha_0 B_0(x)= q_0(x).
\label{bdry}
\eeq

Again, other boundary choices are possible for $A_0(x)$ and $B_0(x)$
as far as (\ref{bdry}) is fullfilled.

Once the expansion is established, any linear combination of distributions has still
to satisfy the same expansion, for appropriate combinations of the scale-invariant
coefficients $A_n(x)$ given above. 

The numerical implementation of this algorithm, especially 
to NLO, is not so straightforward and requires some effort. Among its advantages, however, 
is its remarkable speed. 

\section{Initial conditions and Results}

As input distributions in the unpolarized case, we have used the parametrized
formulas of Ref. \cite{GRV}, that are calculated to NLO in the \( \overline{\textrm{MS}} \)
scheme at a scale \( Q_{0}^{2}=0.40\, \textrm{GeV}^{2} \) \begin{eqnarray}
x(u-\overline{u})(x,Q_{0}^{2}) & = & 0.632x^{0.43}(1-x)^{3.09}(1+18.2x)\nonumber \\
x(d-\overline{d})(x,Q_{0}^{2}) & = & 0.624(1-x)^{1.0}x(u-\overline{u})(x,Q_{0}^{2})\nonumber \\
x(\overline{d}-\overline{u})(x,Q_{0}^{2}) & = & 0.20x^{0.43}(1-x)^{12.4}(1-13.3\sqrt{x}+60.0x)\nonumber \\
x(\overline{u}+\overline{d})(x,Q_{0}^{2}) & = & 1.24x^{0.20}(1-x)^{8.5}(1-2.3\sqrt{x}+5.7x)\nonumber \\
xg(x,Q_{0}^{2}) & = & 20.80x^{1.6}(1-x)^{4.1}
\end{eqnarray}
and \( xq_{i}(x,Q_{0}^{2})=x\overline{q_{i}}(x,Q_{0}^{2})=0 \) for
\( q_{i}=s,c,b,t \).

Following \cite{GRSV}, we have related the unpolarized input distributions
to the longitudinally polarized ones by 

\begin{eqnarray}
x\Delta u(x,Q_{0}^{2}) & = & 1.019x^{0.52}(1-x)^{0.12}xu(x,Q_{0}^{2})\nonumber \\
x\Delta d(x,Q_{0}^{2}) & = & -0.669x^{0.43}xd(x,Q_{0}^{2})\nonumber \\
x\Delta \overline{u}(x,Q_{0}^{2}) & = & -0.272x^{0.38}x\overline{u}(x,Q_{0}^{2})\nonumber \\
x\Delta \overline{d}(x,Q_{0}^{2}) & = & x\Delta \overline{u}(x,Q_{0}^{2})\nonumber \\
x\Delta g(x,Q_{0}^{2}) & = & 1.419x^{1.43}(1-x)^{0.15}xg(x,Q_{0}^{2})
\end{eqnarray}
and \( x\Delta q_{i}(x,Q_{0}^{2})=x\Delta \overline{q_{i}}(x,Q_{0}^{2})=0 \)
for \( q_{i}=s,c,b,t \).

We show in Fig. \ref{uplus} results for the evolution of the $u^+$ 
distribution at the initial scale ($0.632$ GeV) and at two final scales, 
100 GeV and 200 GeV respectively. The peaks at the various scales 
get lowered and become more pronounced toward the smaller x region as Q increases. 
In Fig. \ref{uminus} $u^-$ show an apparent steeper growth at small-x compared to 
$u^+$. For the $d$ distributions the situation is inverted, with $d^-$ 
growing steeper compared to $d^+$ (Figs. \ref{dplus} and \ref{dminus} respectively). 
This apparent behaviour is resolved in Figs. (\ref{ulog}) and (\ref{dlog}) 
from which it is evident that both plus and minus components converge, at very small-x 
values, toward the same limit.   
 
The components s,c,t and b (Figs.~\ref{splus}-\ref{tminus})  
have been generated radiatively from vanishing 
initial conditions for final evolution scales of 100 and 200 GeV $(s,c,b)$ 
and 200 GeV (t). 

Both positive and negative components grow steadily at small-x and are negligible at larger x values. The distribution for the top wuark $(t)$ has been included for
completeness. Given the smaller evolution interval the helicity 
distributions for heavier generations 
are suppressed compared to those of lighter flavours. Gluon helicities 
(Figs. \ref{gplus}, \ref{gminus}) are also enhanced at small-x, and show a 
similar growth. The fine difference between the quark u and d 
distributions are shown in Figs. \ref{ulog} and \ref{dlog}.

Finally in Figs. \ref{ups}, \ref{downs} and \ref{gluons} 
we plot simultaneously longitudinally polarized, unpolarized and 
helicity distributions for up quarks, down quarks and gluons 
at an intermediate factorization scale of 100 GeV, relevant
for experiments at RHIC. Notice that while $u^+$ and $u^-$ are positive 
and their difference $(\Delta u)$ is also positive, for down quarks the two helicity components are positive while their difference $(\Delta d)$ is negative. Gluons, in the model studied 
here, have a positive longitudinal polarization, and their helicity components are also positive. The positive and negative gluon helicities are plotted 
in two separate figures, Figs. \ref{gplus} and \ref{gminus}, while their 
difference, $\Delta g(x)$ is shown in Fig. \ref{Dgluon}. One can observe, at 
least in this model, a crossing at small-x in this distribution. 

We conclude that, at least for this set of boundary conditions, positivity 
of all the components holds to NLO, as expected.  

\section{Conclusions}
We have discussed in detail some of the main features of the 
probabilistic approach to the DGLAP evolution in the helicity basis. 
Numerical results for the evolution of 
all the helicities have been provided, using a 
special algorithm, based in x-space. We have also illustrated some of the 
essential differences between the standard distributional 
form of the kernels and their probabilistic version, claryfying 
some issues connected to their regularization. Then we have turned to 
the probabilistic picture, stressing on the connection between 
the random walk approach to parton diffusion in x-space and the master 
form of the DGLAP equation. The link between the two descriptions 
has been discussed especially in the context of the Kramers-Moyal 
expansion. A Fokker-Planck approximation to the expansion 
has also been presented which may turn useful for the study 
of formal properties of the probabilistic evolution. We have also seen 
that positivity of the helicity distributions, to NLO, 
requires a numerical analysis, as already hinted in \cite{teryaev}.  
Our study also validates the use of a very fast evolution algorithm, 
alternative to other standard algorithms based on Mellin algorithms, 
whose advantage is especially in the analysis of the evolution of nonforward 
parton distributions, as we will show elsewhere.

\section*{Acknowledgements}
We thank O. Teryaev for correspondence and the IFT at the Univ. of Florida for 
hospitality and INFN of Italy (BA-21) for partial financial support.

\appendix

\section{Edge Point Positivity}
We report below 3 tables illustrating the (positive) numerical values of the 
contributions coming from the subtraction terms in the NLO kernels. Coefficients A and B
refer to the subtraction terms $-\log(1- \Lambda ) A + B$ as explained in the section above.

\begin{tabular}{|c|c|c|}
\hline
\( N_{f} \)&
\( A \)&
\( B \)\\
\hline
\hline
3&
12.5302&
12.1739\\
\hline
4&
10.9569&
10.6924\\
\hline
5&
9.3836&
9.2109\\
\hline
6&
7.8103&
7.7249\\
\hline
\end{tabular}
Table 1. Coefficients A and B  for $P^{(1)}_{NS,++}$

\begin{tabular}{|c|c|c|}
\hline
\( N_{f} \)&
\( A \)&
\( B \)\\
\hline
\hline
3&
12.5302&
12.1739\\
\hline
4&
10.9569&
10.6924\\
\hline
5&
9.3836&
9.2109\\
\hline
6&
7.8103&
7.7249\\
\hline
\end{tabular}
Table 2. Coefficients A and B as in Table 1 for $P^{(1)}_{qq,++}$

\begin{tabular}{|c|c|c|}
\hline
\( N_{f} \)&
\( A \)&
\( B \)\\
\hline
\hline
3&
48.4555&
27.3912\\
\hline
4&
45.7889&
24.0579\\
\hline
5&
43.1222&
20.7245\\
\hline
6&
40.4555&
17.3912\\
\hline
\end{tabular}
Table 3. Coefficients A and B as in Table 1 for $P^{(1)}_{gg,++}$

\section{Regularizations}

The ``+'' plus form of the kernels and all the other forms introduced before, 
obtained by separating the 
contributions from the edge-point $(x=1)$ from those coming from the bulk $(0< x < 1)$
  are all equivalent, as we are going to show, 
with the understanding that a linear (unique) cutoff is used to regulate the divergences 
both at x=0 and at x=1. We focus here on the two possible sources of singularity, i.e. on 
$P_{qq}$ and on the $P_{gg}$ contributions, which require some attention. 
Let's start from the $P_{qq}$ case.   
We recall that ``+'' plus distributions are defined as  

\beq
\frac{1}{(1-x)_+}=\frac{\theta(1- x- \Lambda)}{1-x} -
\delta(1-x) \int_0^{1-\Lambda}\frac{dz}{1-z}
\eeq
with $\Lambda$ being a cutoff for the edge-point contribution. 

We will be using the relations 
\beqa
\int_x^1\frac{dy}{y}\delta(1-y) &=& 1 \nonumber \\
\int_0^1\frac{dz}{1-z}&=&\int_x^1\frac{dz}{1-z} - log(1-x)\nonumber \\
\int_x^1\frac{dy}{y} f(y)g(x/y)&=& \int_x^1\frac{dy}{y} f(x/y)g(y).
\label{refs}
\eeqa

Using the expressions above it is easy to obtain 
\beqa
\frac{1}{(1-x)_+}\otimes f(x)&=&\int_x^1\frac{dy}{y}\frac{1}{(1- x/y)_+} q(y) 
\nonumber \\
&=&\int_x^1\frac{dy}{y}\frac{1}{1- x/y}f(y) -\int_x^1\frac{dy}{y}
\delta(1-y)\int_0^1 \frac{dz}{z} \nonumber \\
&=&\int_x^1\frac{dy}{y}
\frac{ yf(y) - x f(x)}{y-x} +\log(1-x) f(x)\nonumber \\
\eeqa
which is eq. (\ref{simplerel}). If we remove the ``+'' distributions 
and adopt (implicitely) a cutoff regularization, we need special care. 
In the probabilistic version of the kernel, the handling of 
$P_{qq}\otimes q$ is rather straightforward 

\beqa
P_{qq}\otimes q(x) &=& C_F\int_x^1\frac{dy}{y}
\left( \frac{2}{1- x/y} -1- x/y\right)q(y)\nonumber \\
&& - C_F\int_x^1\frac{dy}{y}\delta(1-y)\int_0^1 dz'
\left( \frac{2}{1- z'} -1- z'\right) \nonumber \\
\eeqa
and using eqs. (\ref{refs}) we easily obtain 
\beqa
P_{qq}\otimes q(x) &=& 2 C_F \int_x^1\frac{dy}{y}
\frac{y q(y) - x q(x)}{y-x} + 2 C_F \log(1-x) q(x) \nonumber \\
 && - C_F\int_x^1\frac{dy}{y}\left( 1 + x/y\right)q(y) + \frac{3}{2}C_F q(x).
\eeqa
Now consider the convolution $P_{gg}\otimes g(x)$ in the Collins-Qiu form. 
We get 
\beqa 
P_{gg}\otimes g(x)&=& 2 C_A \int_x^1\frac{dy}{y}\frac{1}{1- x/y}g(y) 
\nonumber \\
&& + 2 C_A \int_x^1\frac{dy}{y}\left( x/y (1 - x/y) + \frac{1}{x/y} -2 
\right)g(y)
- g(x)\int_0^1\frac{dz} C_A \left( \frac{1}{z} 
+ \frac{1}{1-z}\right)
\nonumber \\
&& -\frac{1}{2}g(x)\int_0^1 dz \,\,2 C_A\left( z(1-z) - 2\right)
- n_f g(x)\int_0^1 dz \frac{1}{2}\left( z^2 + (1-z)^2\right).
\eeqa

There are some terms in the expression above that require some care. 
The appropriate regularization is 
\beq
\int_0^1\frac{dz}{z} + \int_0^1 \frac{dz}{1-z}\rightarrow 
I(\Lambda)= \int_\Lambda^1\frac{dz}{z} + \int_0^{1- \Lambda} \frac{dz}{1-z}.
\eeq

Observe also that 
\beq
 \int_\Lambda^1\frac{dz}{z} = \int_0^{1- \Lambda} \frac{dz}{1-z}= -\log \Lambda.
\eeq
Notice that in this regularization the singularity of $1/z$ at $z=0$ 
is traded for a singularity at z=1 in $1/(1-z)$. 
 It is then rather straightforward to show that 
\beqa
P_{gg}\otimes g(x) &=& 2 C_A 
\int_x^1\frac{dy}{y}
\frac{ yg(y) - x g(x)}{y-x} + 2 C_A\log(1-x) g(x) \nonumber \\
&& + 2 C_A \int_x^1\frac{dy}{y}\left( x/y ( 1 - x/y) 
+ \frac{1}{x/y} - 2 \right)
+ \frac{\beta_0}{2} g(x).
\eeqa

A final comment is due for the form of the sum rule 
\beq
\int_0^1 dz (z - \frac{1}{2})\gamma_{gg}=0
\label{gluonsr}
\eeq
that we need to check with the regularization given above.

The strategy to handle this expression is the same as before. 
We extract  all the $1/z$ and $1/(1-z)$ integration terms and use 
\beqa
\int_0^1 \frac{dz}{z} - \int_0^1\frac{dz}{1-z}
&\rightarrow & \int_\Lambda^1 \frac{dz}{z} - \int_0^{1- \Lambda}\frac{dz}{1-z}
\nonumber \\
&=& -\log \Lambda + \log \Lambda = 0 \nonumber \\
\eeqa
to eliminate the singularities at the boundaries 
$x=0, 1$ and verify eq. (\ref{gluonsr}).

\begin{figure}[tbh]
{\centering \resizebox*{12cm}{!}{\rotatebox{-90}{\includegraphics{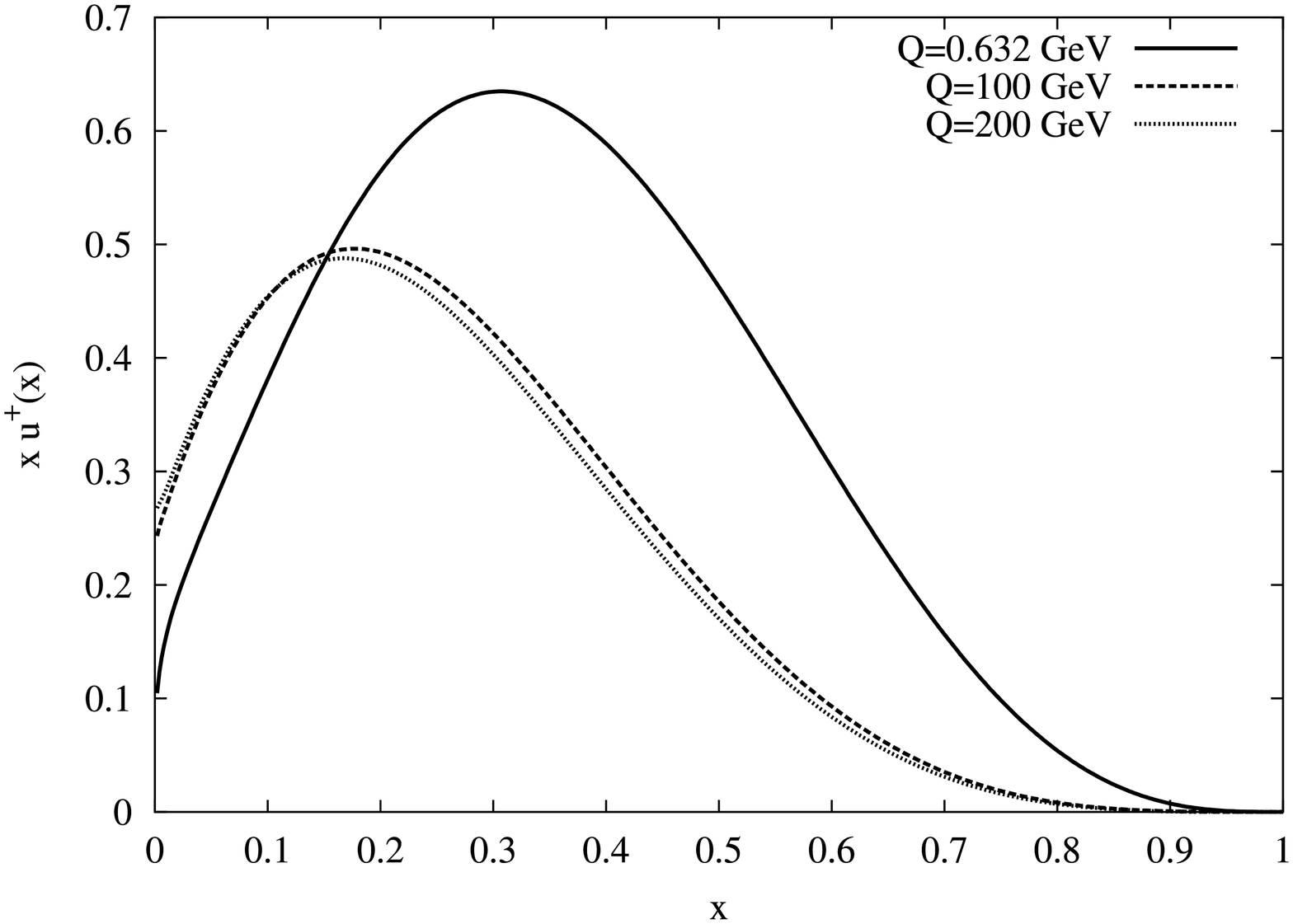}}} \par}

\caption{Evolution of \protect\( u^{+}\protect \) versus \protect\( x\protect \)
at various \protect\( Q\protect \) values.}
\label{uplus}
\end{figure}

\begin{figure}[tbh]
{\centering \resizebox*{12cm}{!}{\rotatebox{-90}{\includegraphics{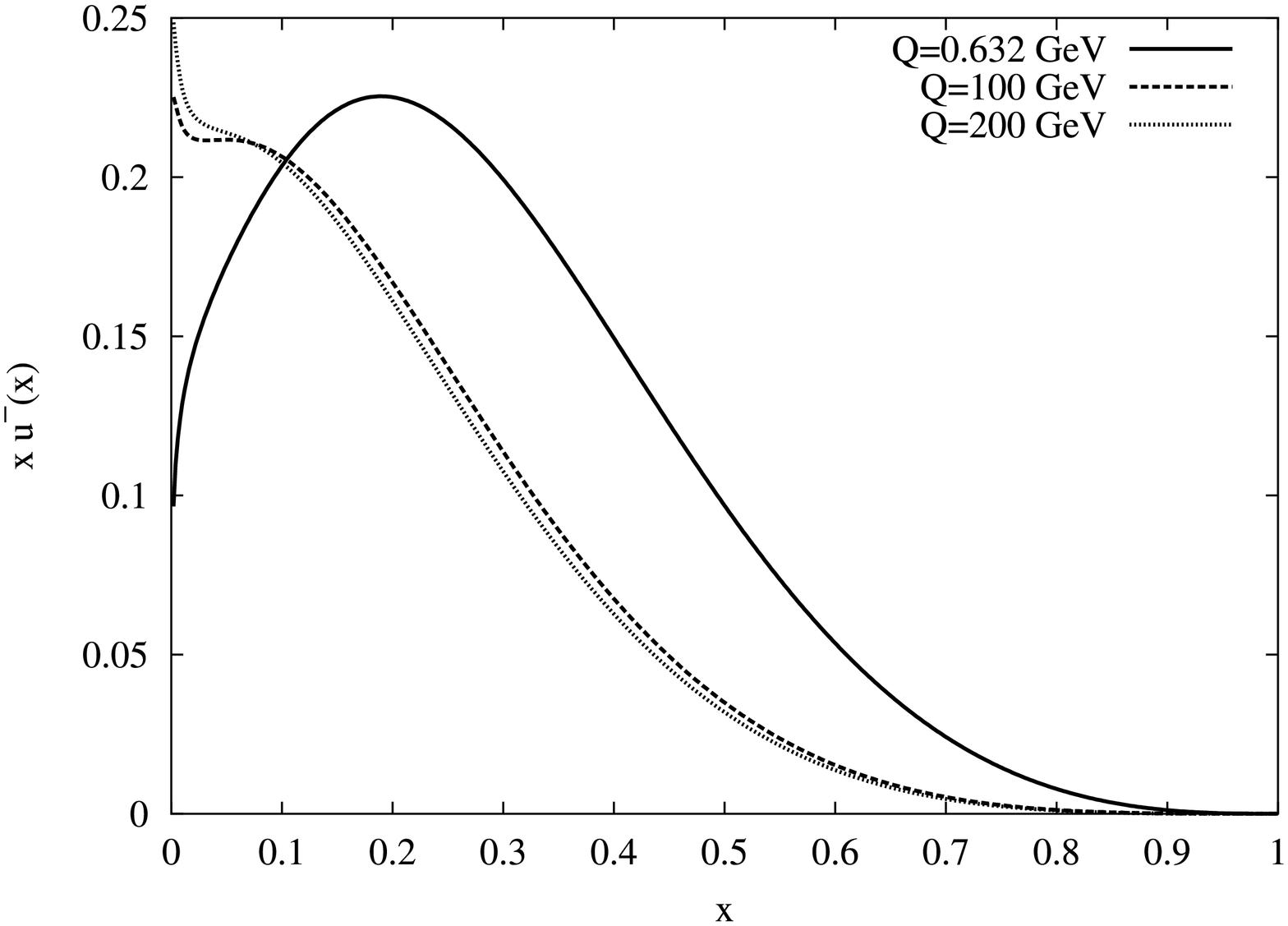}}} \par}

\caption{Evolution of \protect\( u^{-}\protect \) versus \protect\( x\protect \)
at various \protect\( Q\protect \) values.}
\label{uminus}
\end{figure}

\begin{figure}[tbh]
{\centering \resizebox*{12cm}{!}{\rotatebox{-90}{\includegraphics{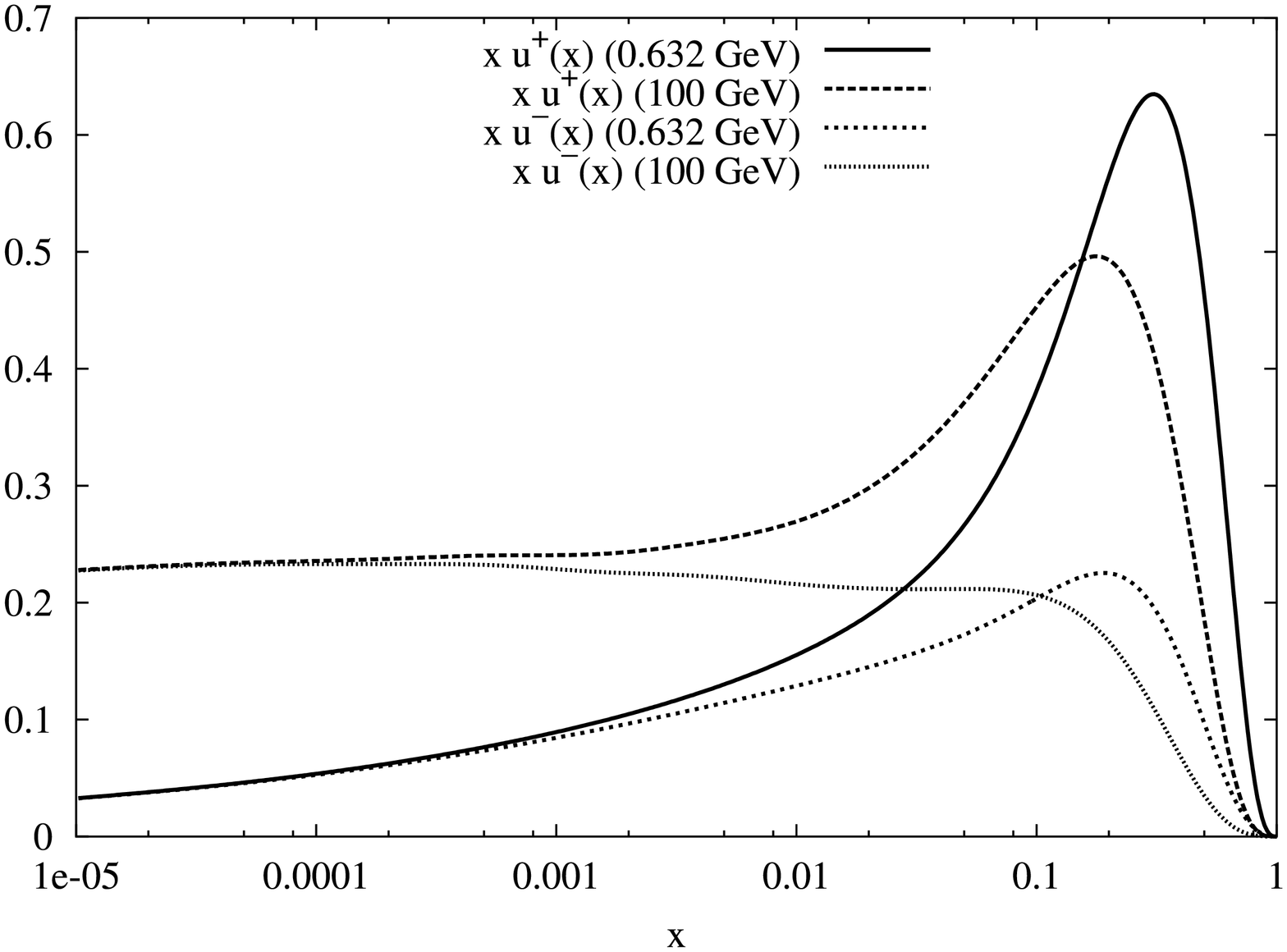}}} \par}
\caption{Small-$x$ behaviour of $u^{\pm}$ at 100 GeV.}
\label{ulog}
\end{figure}

\begin{figure}[tbh]
{\centering \resizebox*{12cm}{!}{\rotatebox{-90}{\includegraphics{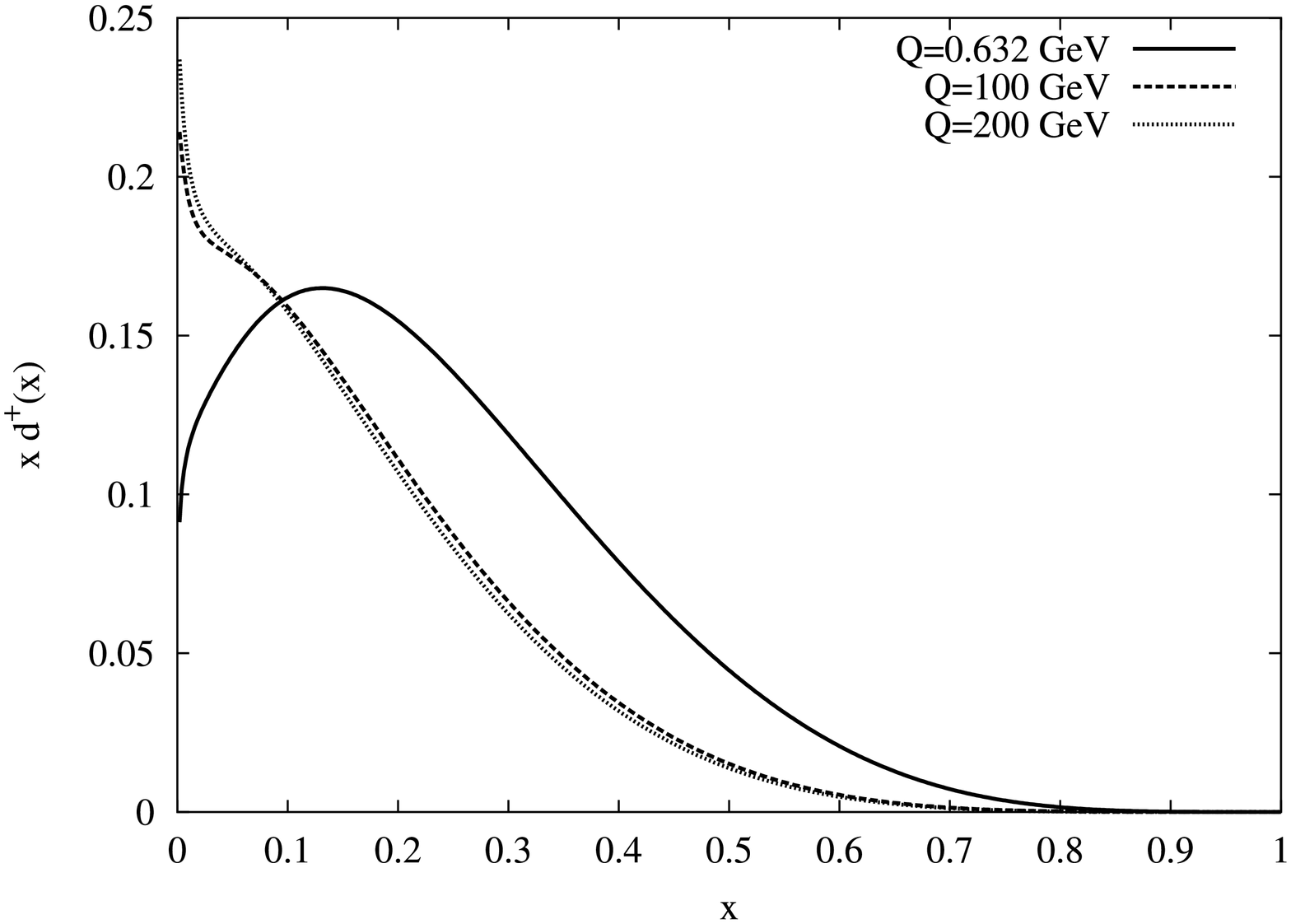}}} \par}
\caption{Evolution of \protect\( d^{+}\protect \) versus \protect\( x\protect \)
at various \protect\( Q\protect \) values.}
\label{dplus}
\end{figure}

\begin{figure}[tbh]
{\centering \resizebox*{12cm}{!}{\rotatebox{-90}{\includegraphics{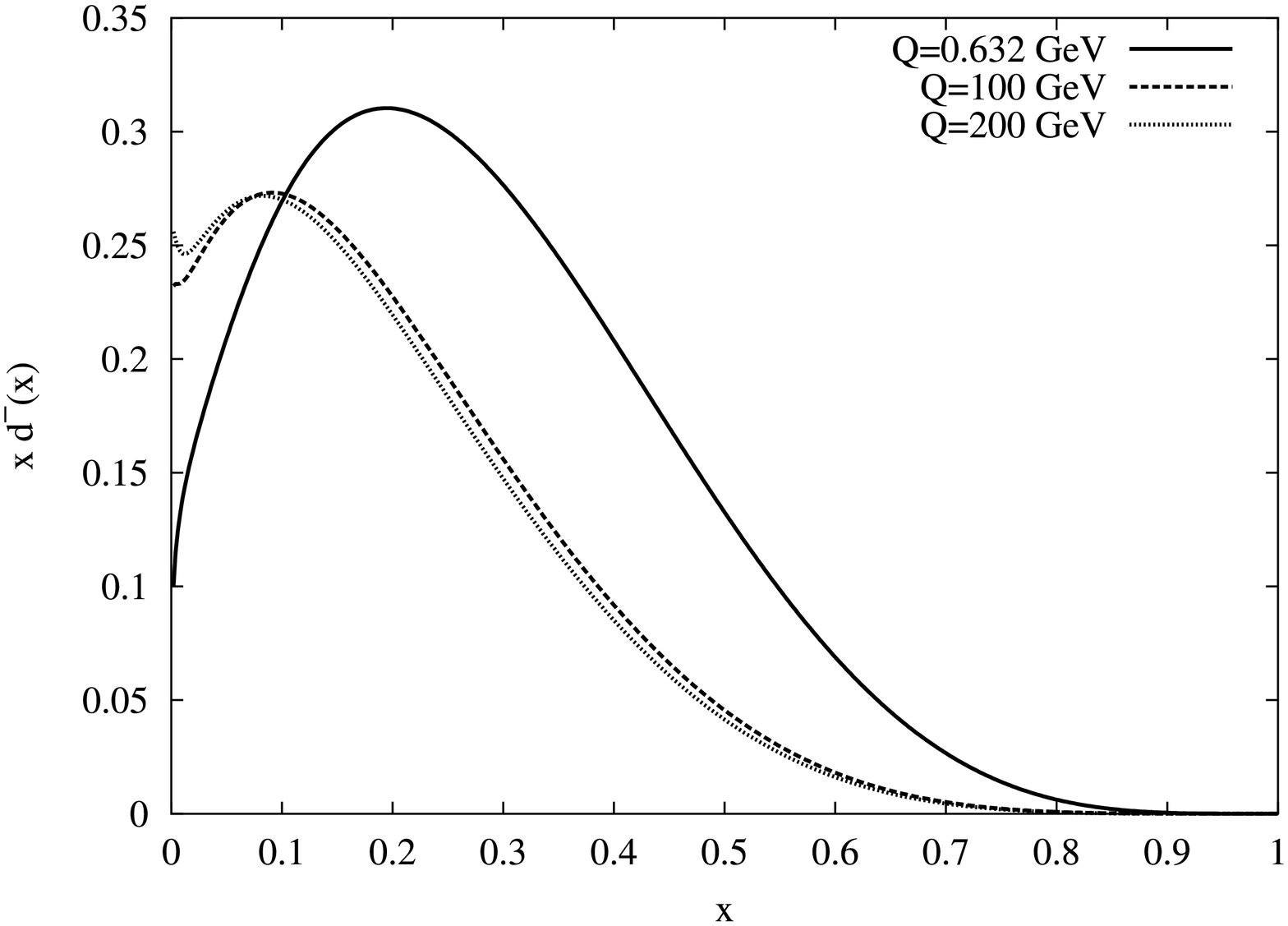}}} \par}
\caption{Evolution of \protect\( d^{-}\protect \) versus \protect\( x\protect \)
at various \protect\( Q\protect \) values.}
\label{dminus}
\end{figure}

\begin{figure}[tbh]
{\centering \resizebox*{12cm}{!}{\rotatebox{-90}{\includegraphics{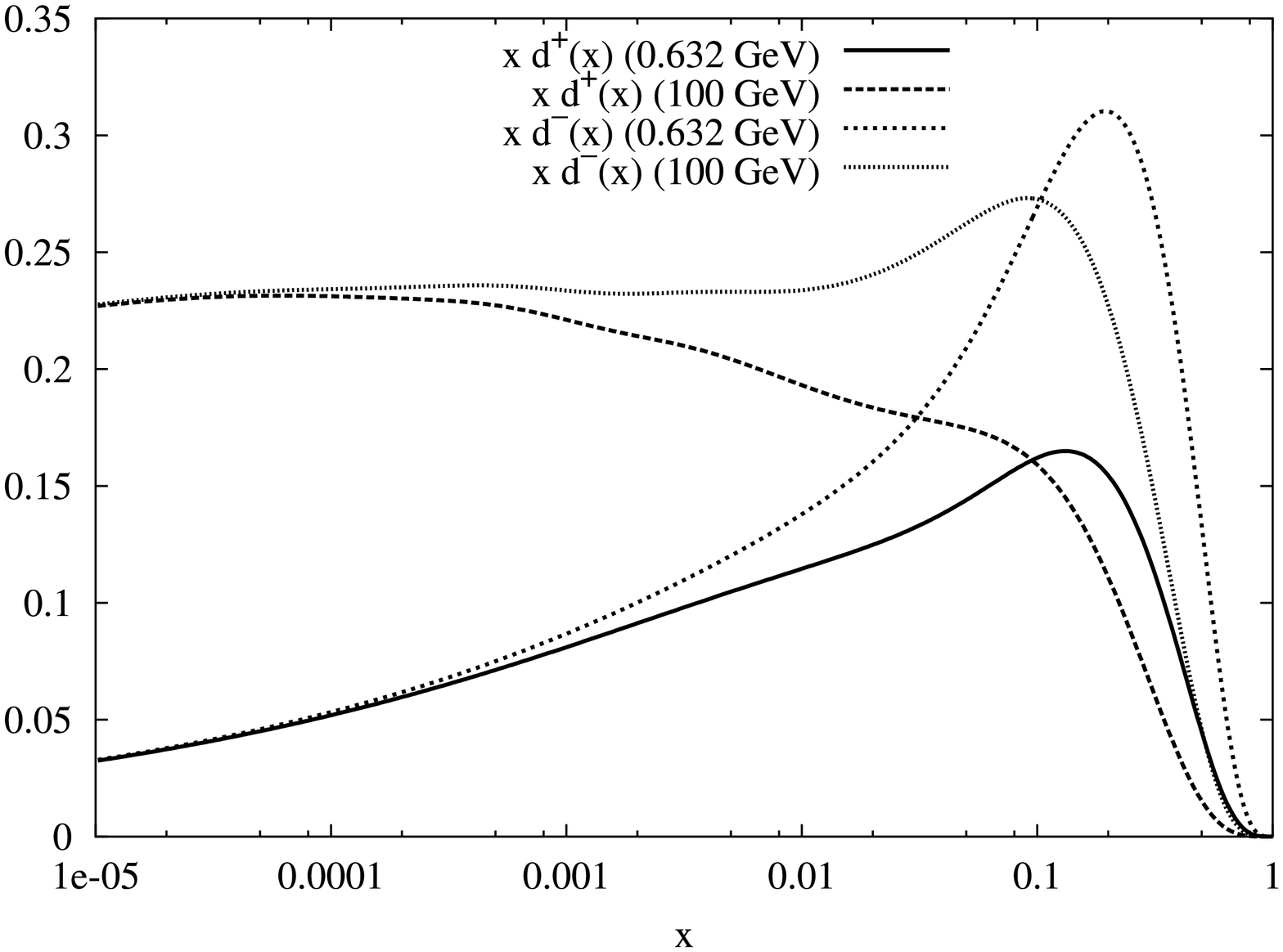}}} \par}
\caption{Small-$x$ behaviour of $d^{\pm}$ at 100 GeV.}
\label{dlog}
\end{figure}

\begin{figure}[tbh]
{\centering \resizebox*{12cm}{!}{\rotatebox{-90}{\includegraphics{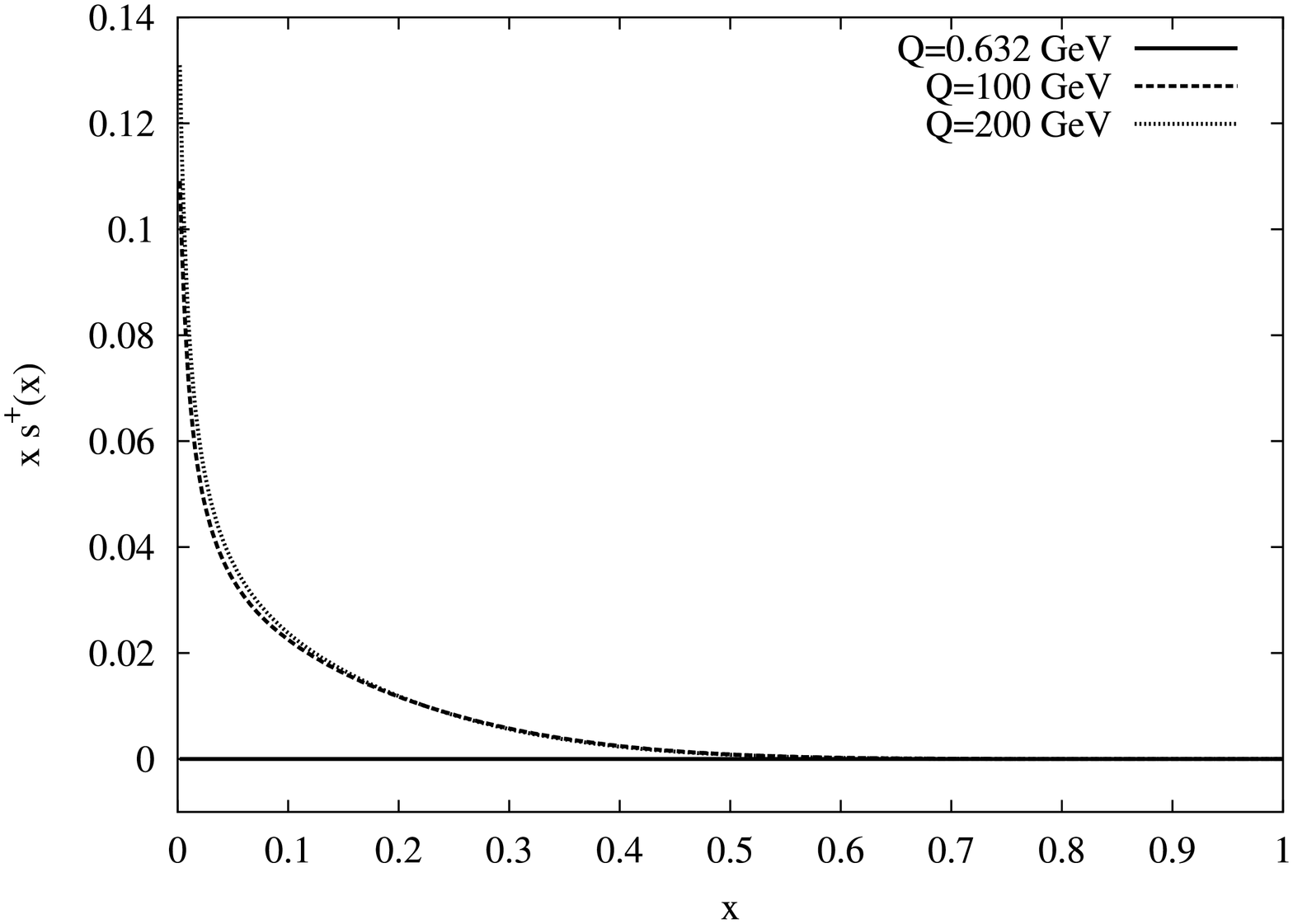}}} \par}

\caption{Evolution of \protect\( s^{+}\protect \) versus \protect\( x\protect \)
at various \protect\( Q\protect \) values.}
\label{splus}
\end{figure}

\begin{figure}[tbh]
{\centering \resizebox*{12cm}{!}{\rotatebox{-90}{\includegraphics{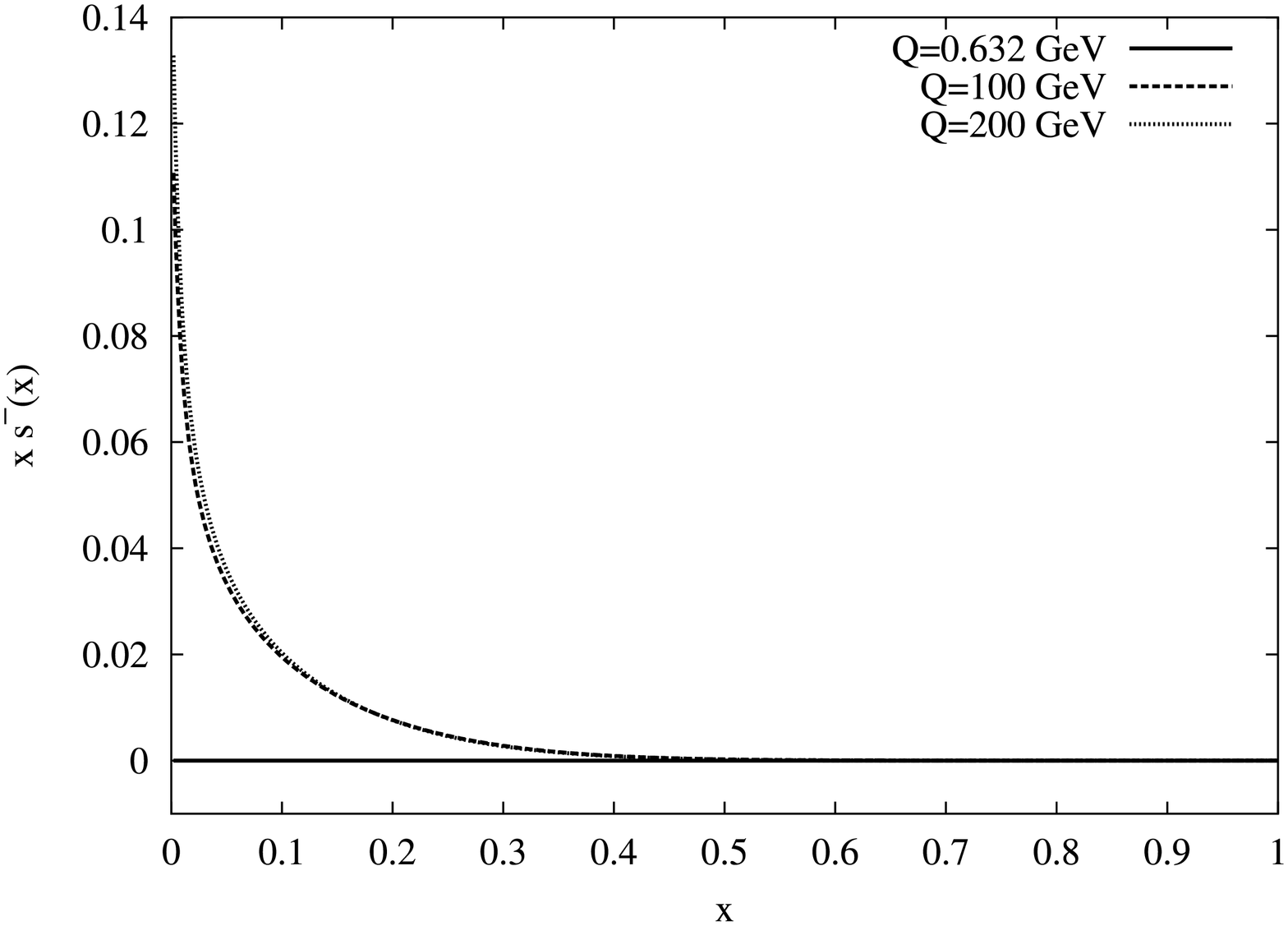}}} \par}

\caption{Evolution of \protect\( s^{-}\protect \) versus \protect\( x\protect \)
at various \protect\( Q\protect \) values.}
\label{sminus}
\end{figure}

\begin{figure}[tbh]
{\centering \resizebox*{12cm}{!}{\rotatebox{-90}{\includegraphics{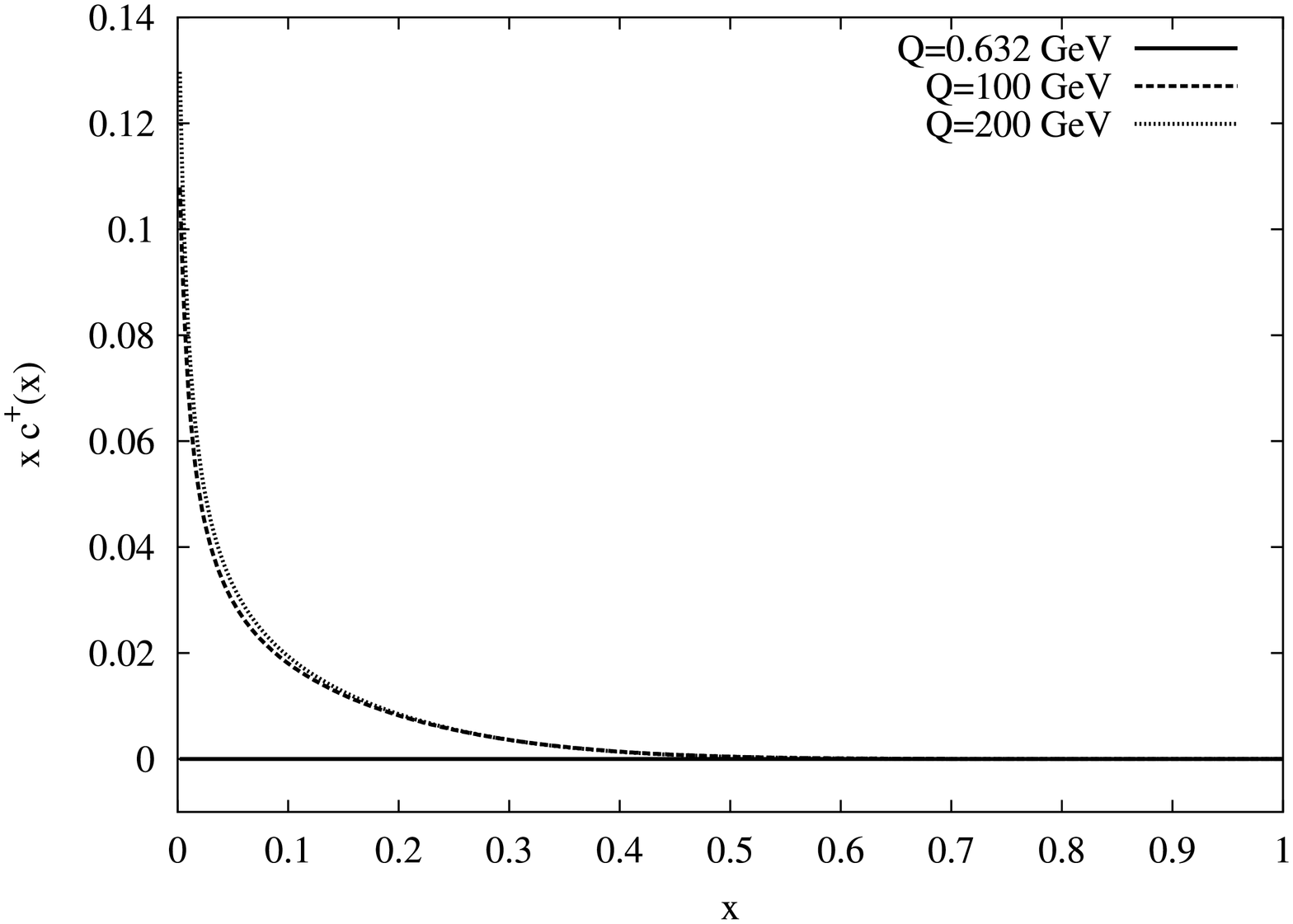}}} \par}

\caption{Evolution of \protect\( c^{+}\protect \) versus \protect\( x\protect \)
at various \protect\( Q\protect \) values.}
\label{cplus}
\end{figure}

\begin{figure}[tbh]
{\centering \resizebox*{12cm}{!}{\rotatebox{-90}{\includegraphics{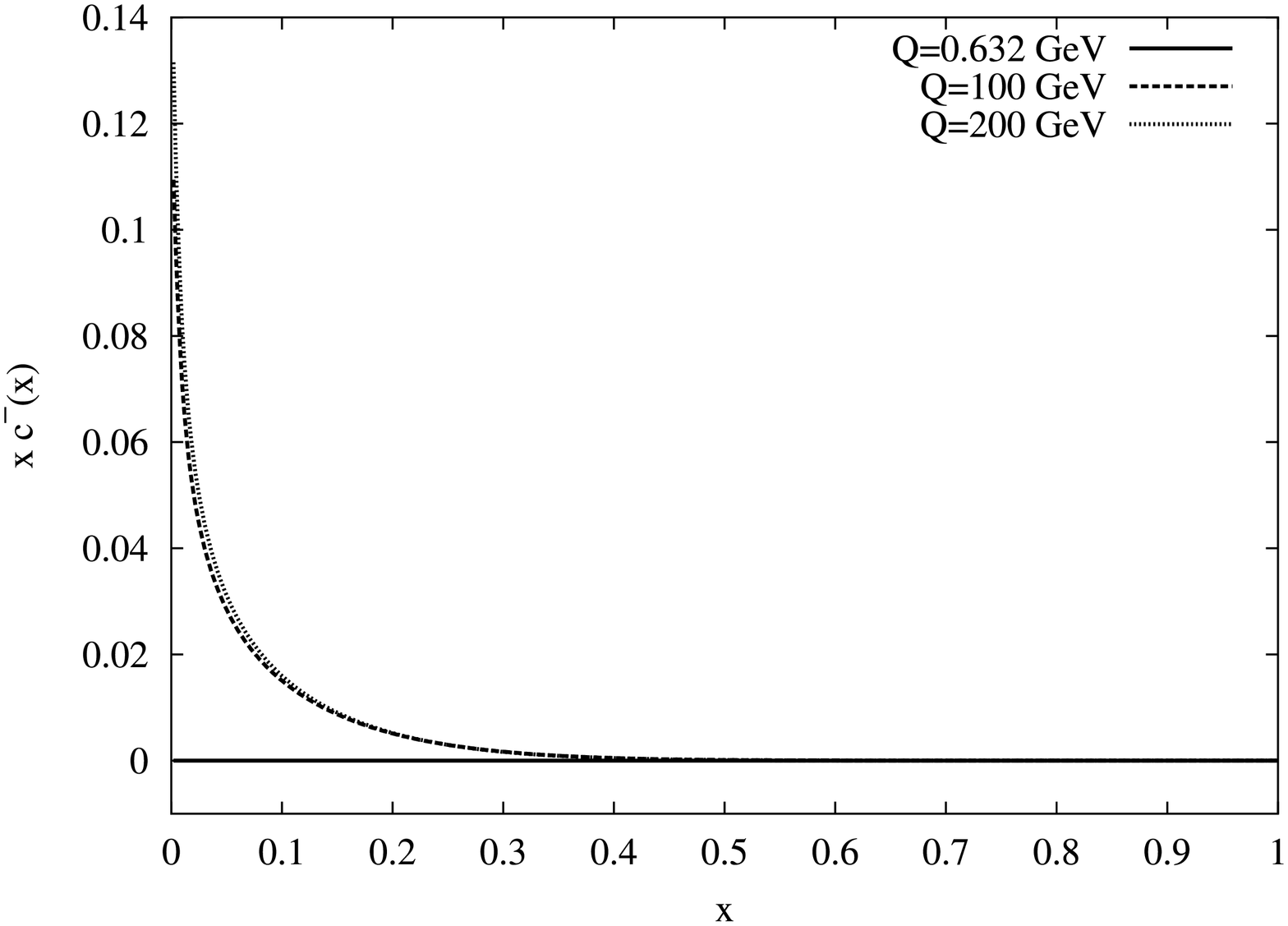}}} \par}

\caption{Evolution of \protect\( c^{-}\protect \) versus \protect\( x\protect \)
at various \protect\( Q\protect \) values.}
\label{cminus}
\end{figure}

\begin{figure}[tbh]
{\centering \resizebox*{12cm}{!}{\rotatebox{-90}{\includegraphics{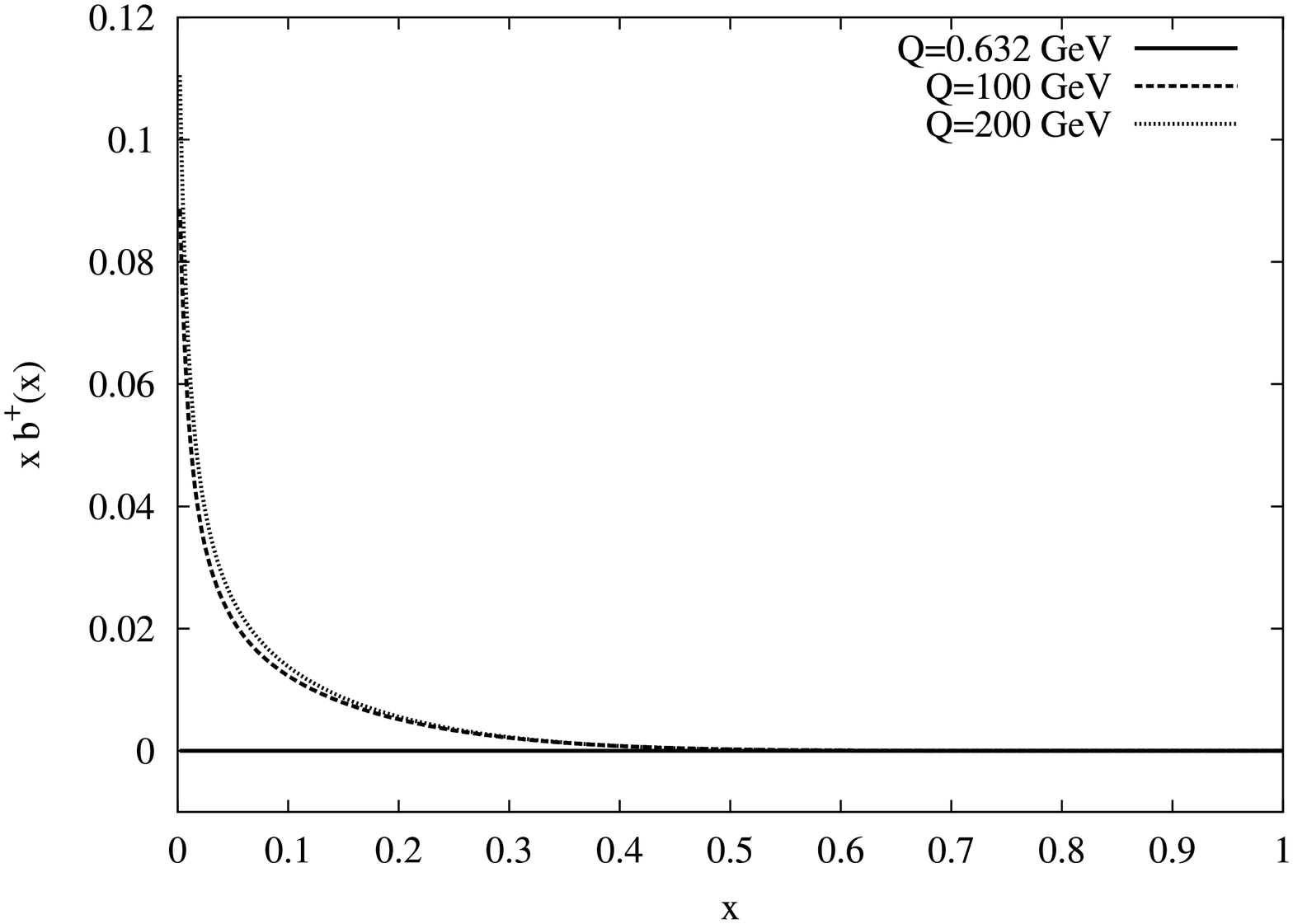}}} \par}

\caption{Evolution of \protect\( b^{+}\protect \) versus \protect\( x\protect \)
at various \protect\( Q\protect \) values.}
\label{bplus}
\end{figure}

\clearpage

\begin{figure}[tbh]
{\centering \resizebox*{12cm}{!}{\rotatebox{-90}{\includegraphics{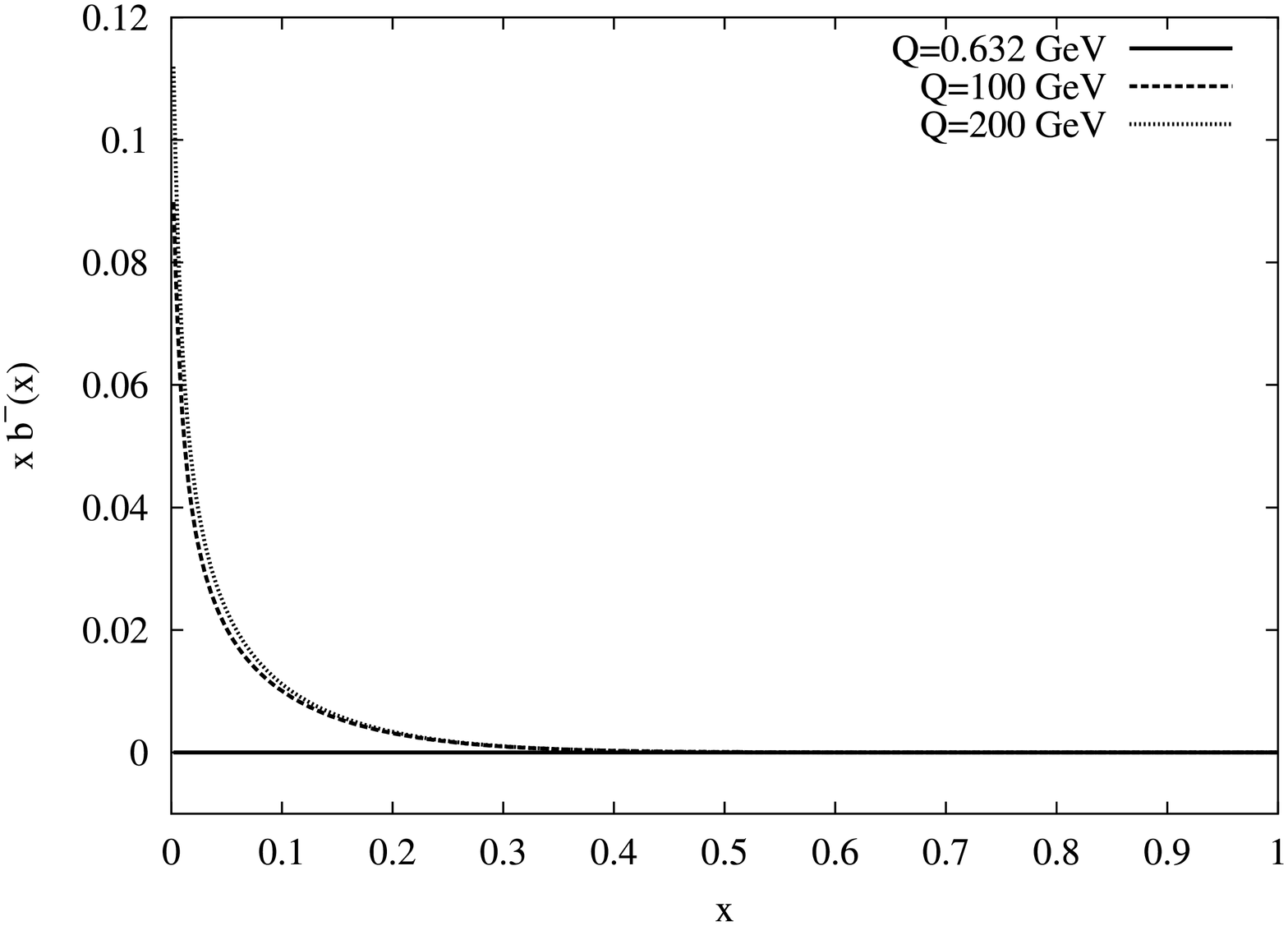}}} \par}

\caption{Evolution of \protect\( b^{-}\protect \) versus \protect\( x\protect \)
at various \protect\( Q\protect \) values.}
\label{bminus}
\end{figure}

\begin{figure}[tbh]
{\centering \resizebox*{12cm}{!}{\rotatebox{-90}{\includegraphics{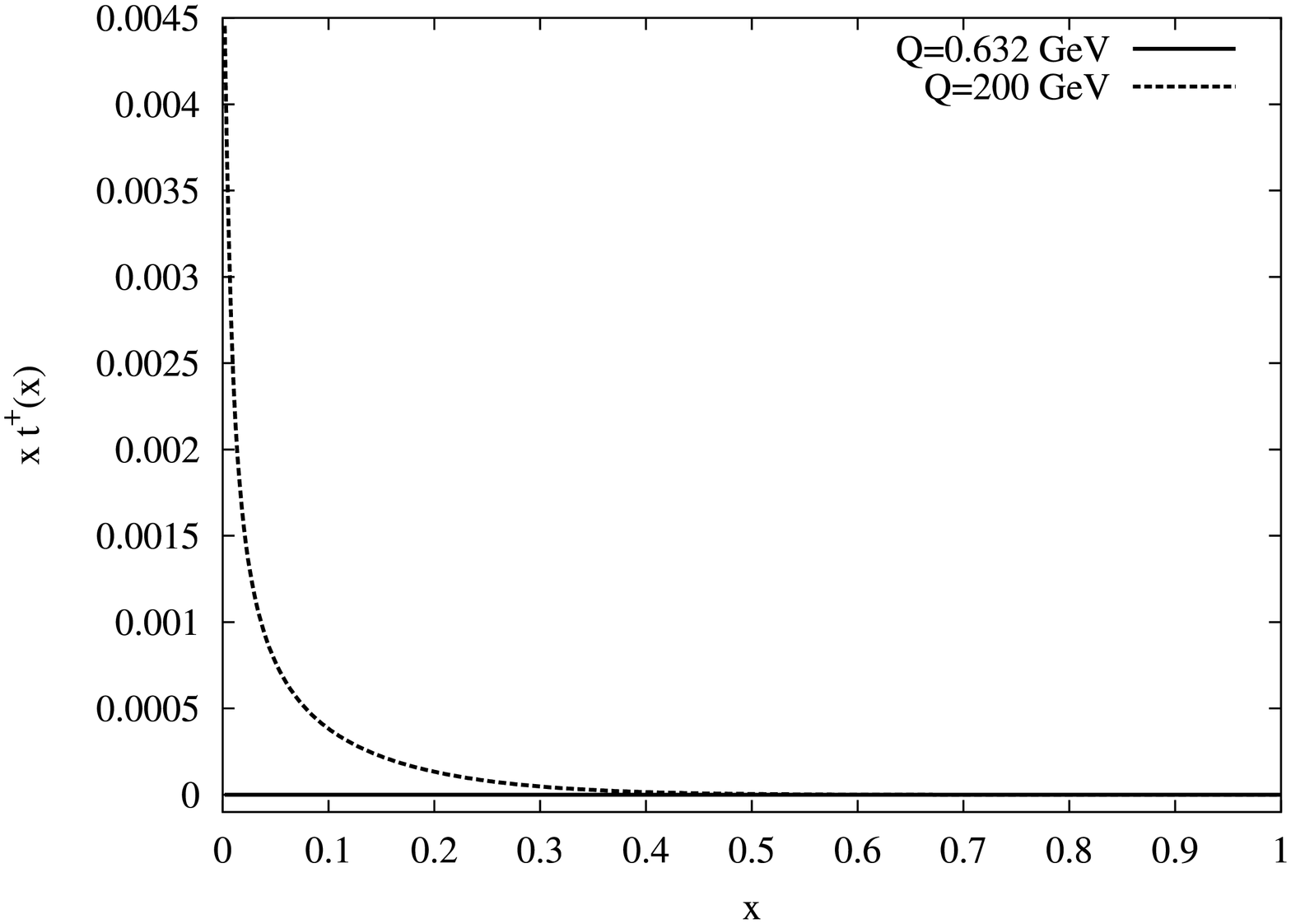}}} \par}

\caption{Evolution of \protect\( t^{+}\protect \) versus \protect\( x\protect \)
at various \protect\( Q\protect \) values.}
\label{tplus}
\end{figure}

\begin{figure}[tbh]
{\centering \resizebox*{12cm}{!}{\rotatebox{-90}{\includegraphics{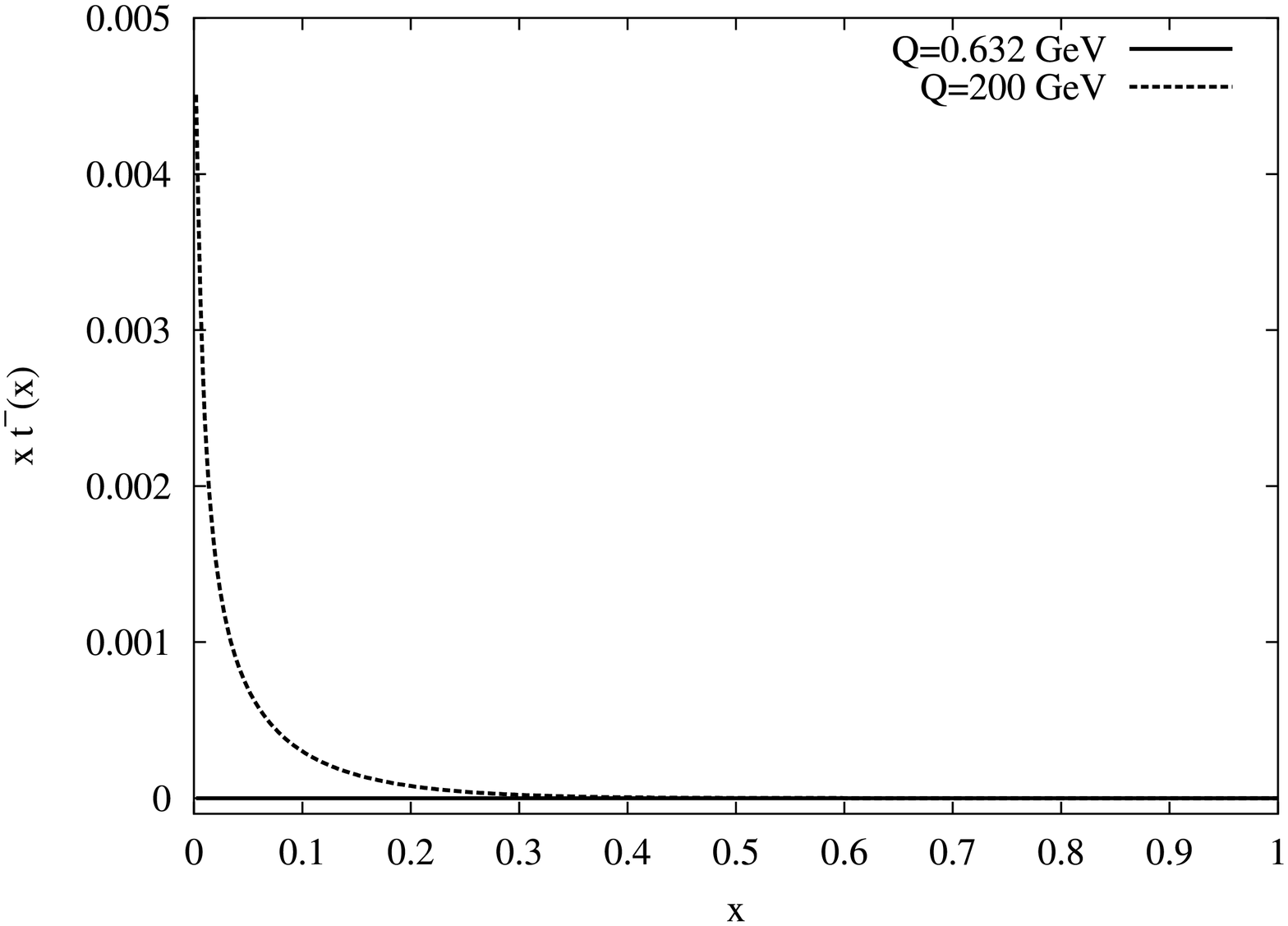}}} \par}

\caption{Evolution of \protect\( t^{-}\protect \) versus \protect\( x\protect \)
at various \protect\( Q\protect \) values.}
\label{tminus}
\end{figure}

\begin{figure}[tbh]

{\centering \resizebox*{12cm}{!}{\rotatebox{-90}{\includegraphics{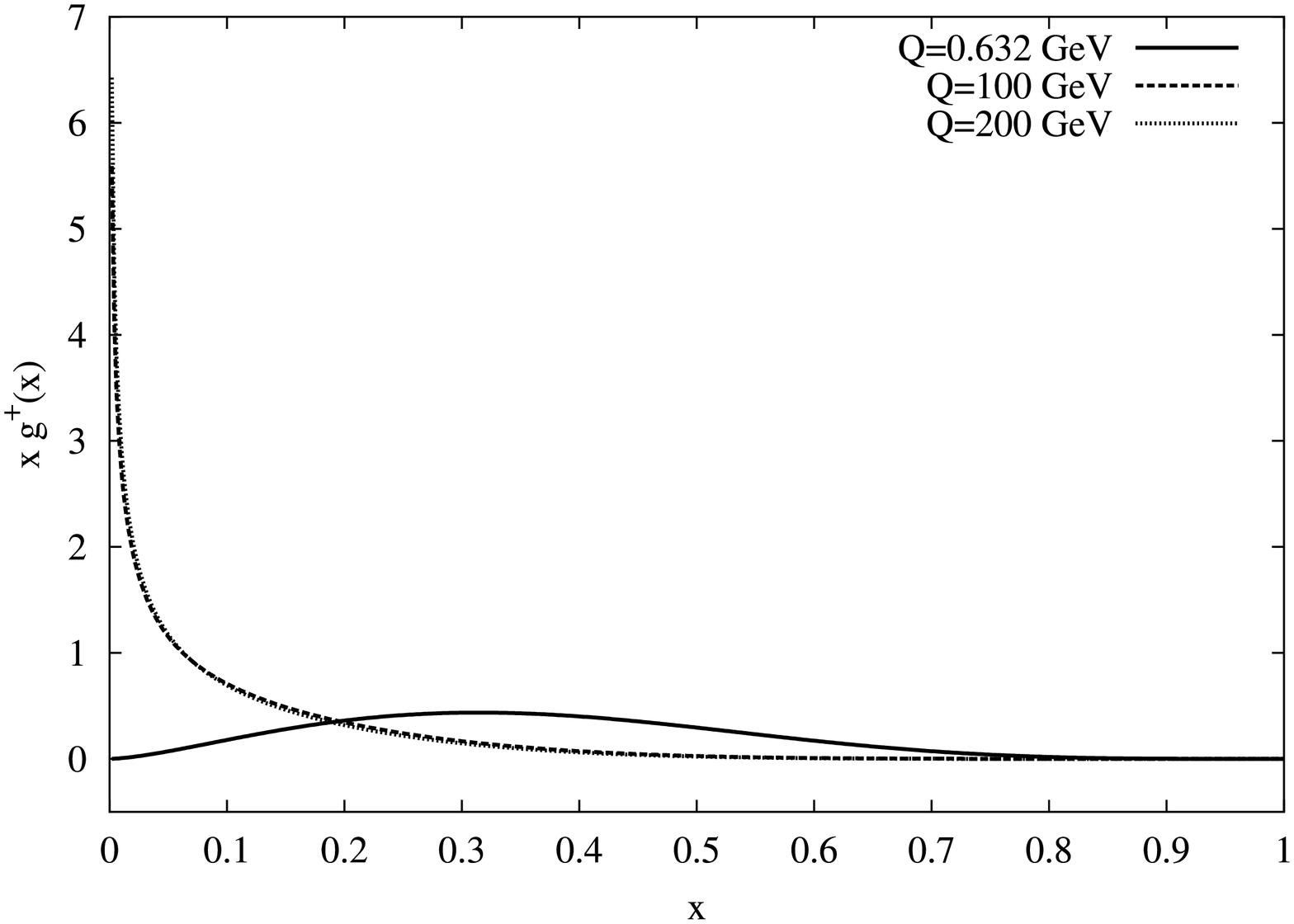}}} \par}

\caption{Evolution of \protect\( g^{+}\protect \) versus \protect\( x\protect \)
at various \protect\( Q\protect \) values.}
\label{gplus}
\end{figure}

\begin{figure}[tbh]
{\centering \resizebox*{12cm}{!}{\rotatebox{-90}{\includegraphics{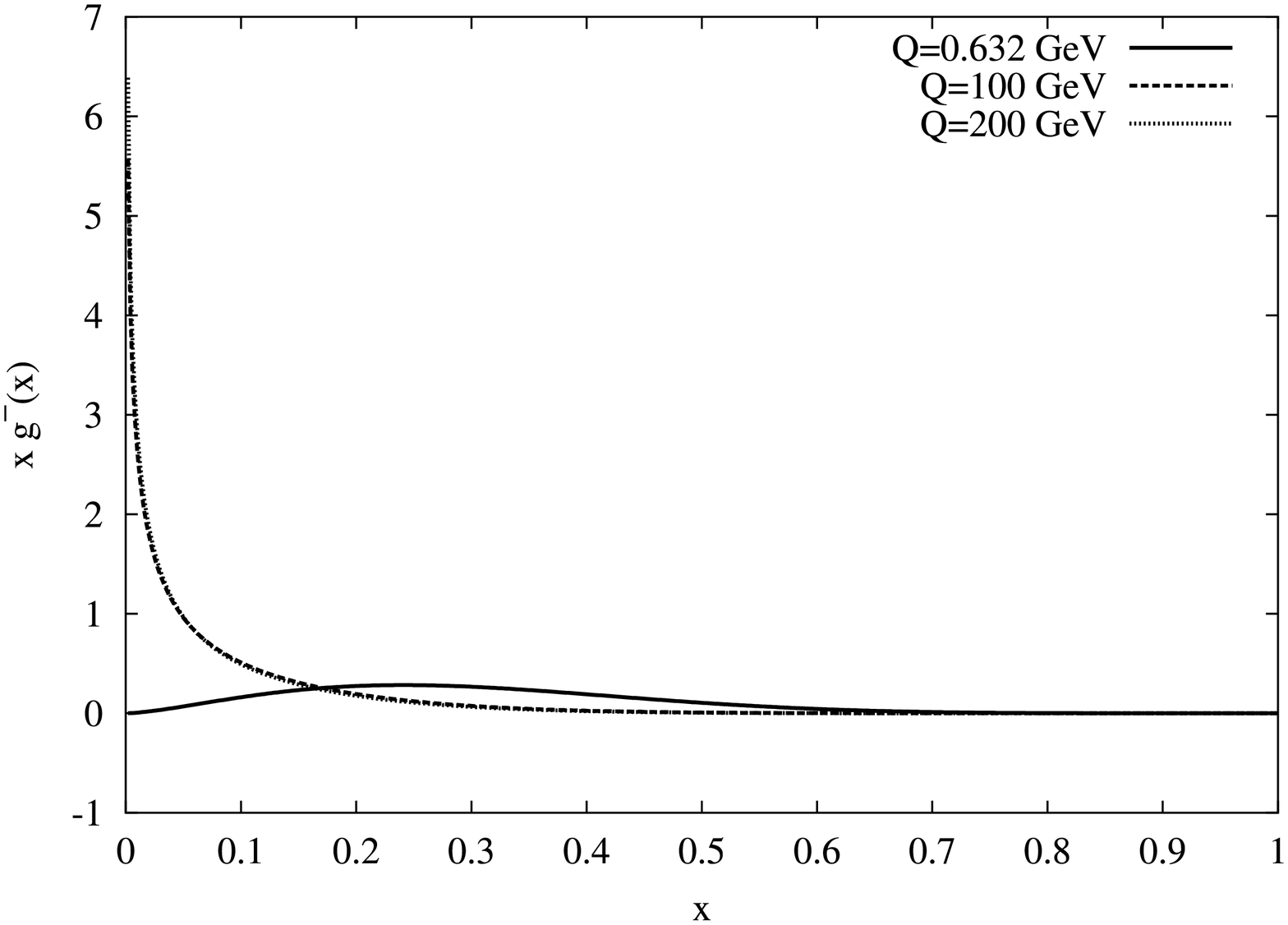}}} \par}

\caption{Evolution of \protect\( g^{-}\protect \) versus \protect\( x\protect \)
at various \protect\( Q\protect \) values.}
\label{gminus}
\end{figure}

\begin{figure}[tbh]
{\centering \resizebox*{12cm}{!}{\rotatebox{-90}{\includegraphics{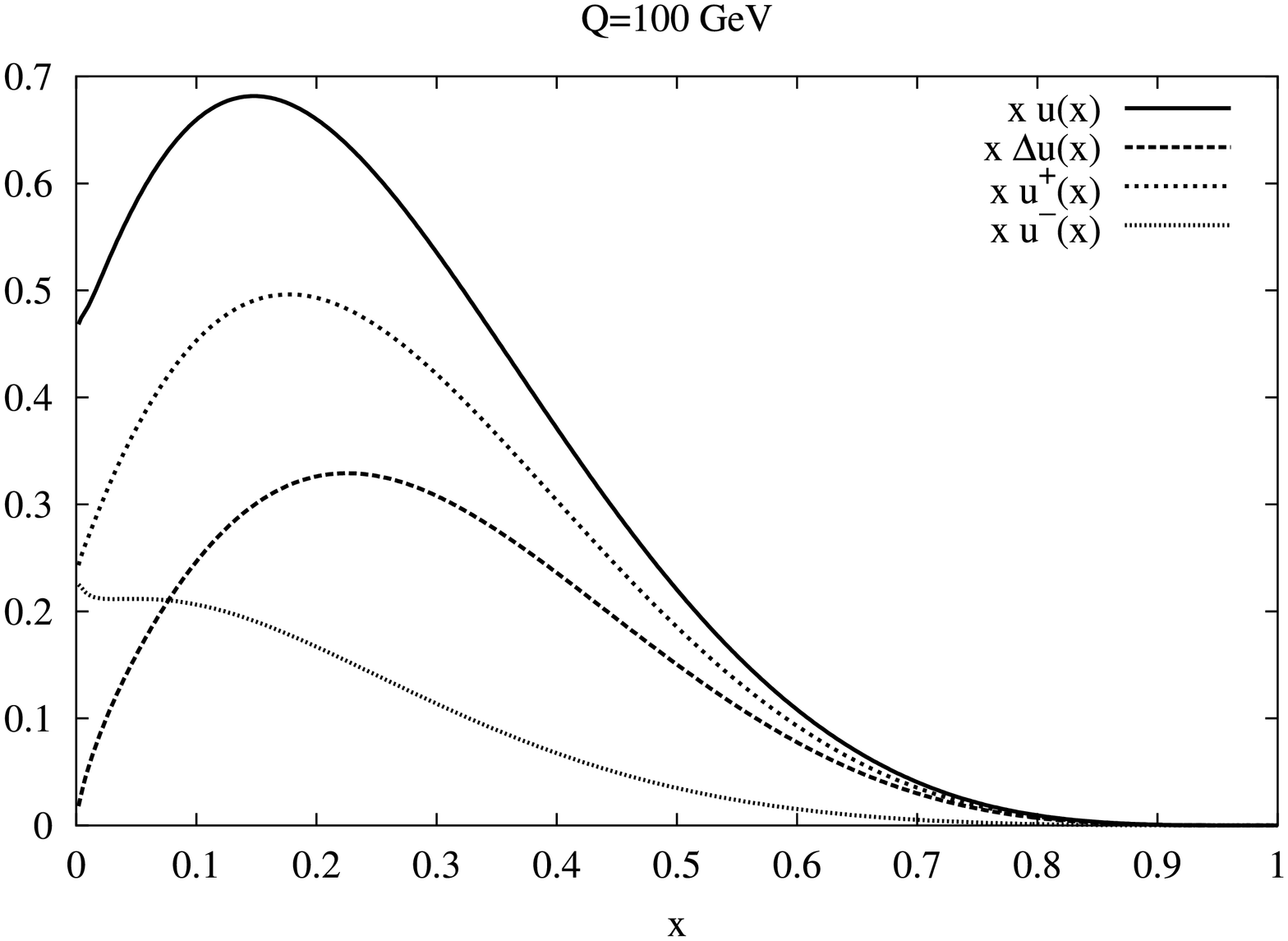}}} \par}
\caption{Various kinds of distributions of quark up at \protect\( Q=100\, \textrm{GeV}\protect \).}
\label{ups}
\end{figure}

\begin{figure}[tbh]
{\centering \resizebox*{12cm}{!}{\rotatebox{-90}{\includegraphics{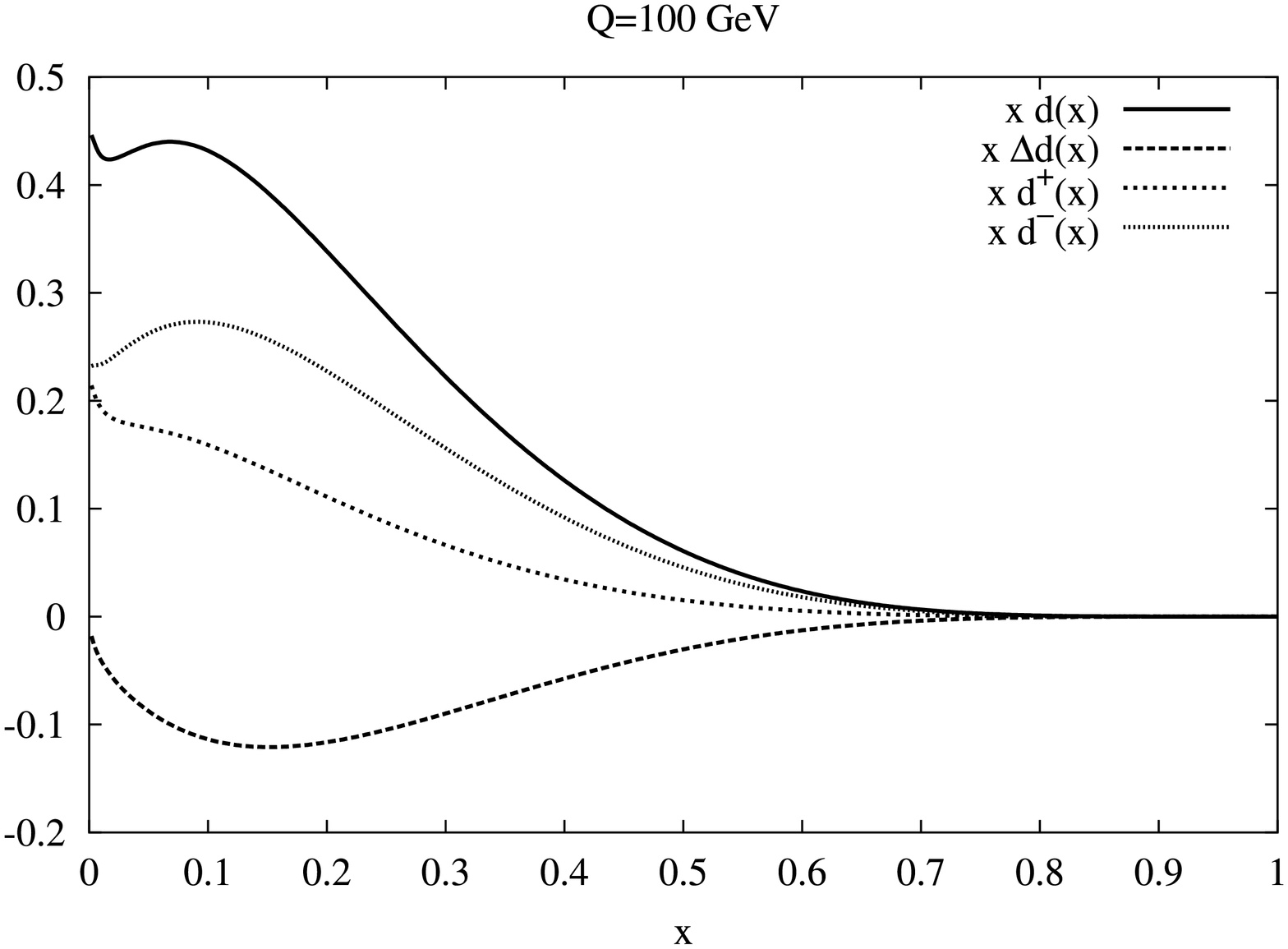}}} \par}
\caption{Various kinds of distributions of quark down at \protect\( Q=100\, \textrm{GeV}\protect \).}
\label{downs}
\end{figure}

\begin{figure}[tbh]
{\centering \resizebox*{12cm}{!}{\rotatebox{-90}{\includegraphics{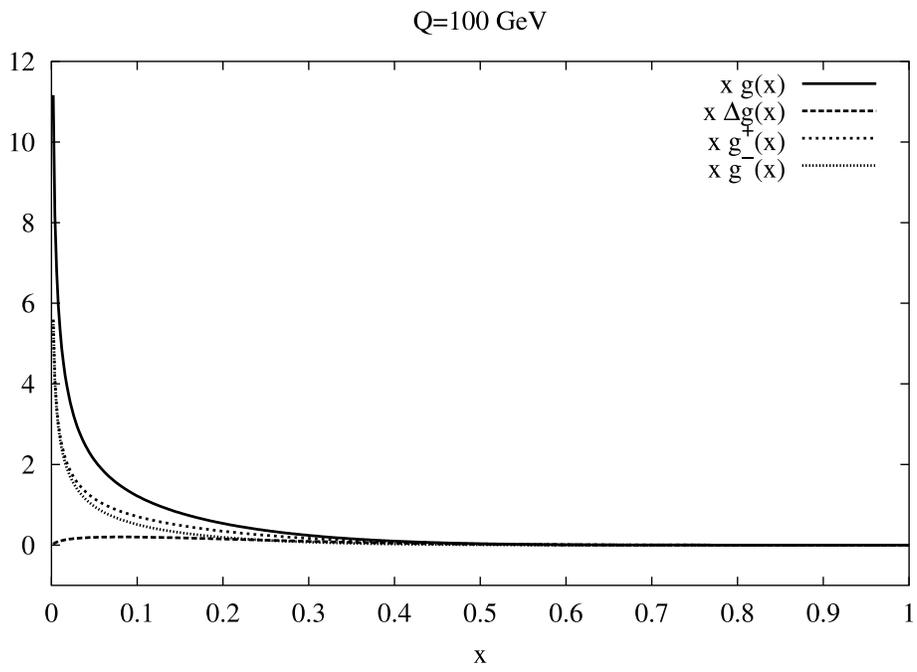}}} \par}

\caption{Various kinds of gluon distributions at \protect\( Q=100\, \textrm{GeV}\protect \).}
\label{gluons}
\end{figure}

\begin{figure}[tbh]
{\centering \resizebox*{12cm}{!}{\rotatebox{-90}{\includegraphics{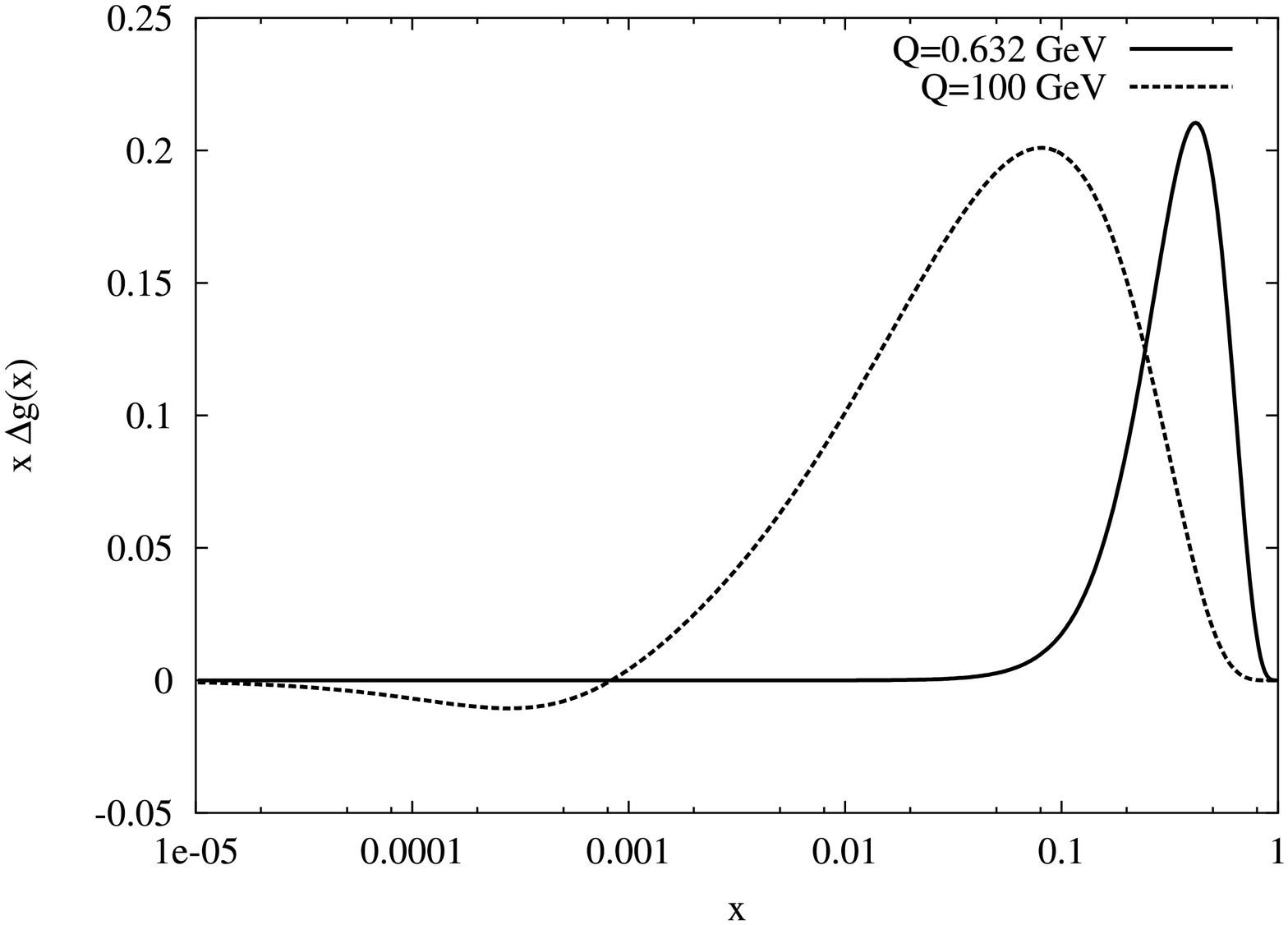}}} \par}

\caption{Small-x behaviour of $\Delta g$ at 100 GeV.}
\label{Dgluon}
\end{figure}

\section{Kernels in the helicity basis}
The expression of the kernels in the helicity basis given below are 
obtained combining the NLO computations of \cite{MVN,CFP,Vog}

\begin{eqnarray}
P^{(1)}_{NS^{-},++}(x) & = & \left\{ \frac{C_{F}}{18}\left[ 90C_{F}(x-1)+4T_{f}(11x-1)+N_{C}(53-187x+3\pi ^{2}(1+x))\right] \right\} \nonumber \\
 &  & +\left\{ \frac{C_{F}\left[ 6C_{F}(3-2(x-1)x)+4T_{f}(1+x^{2})-N_{C}(17+5x^{2})\right] }{6(x-1)}\right\} \log x\nonumber \\
 &  & +\left\{ \frac{C_{F}\left[ C_{F}-N_{C}-(C_{F}+N_{C})x^{2}\right] }{2(x-1)}\right\} \log ^{2}x\nonumber \\
 &  & +\left\{ \frac{2C_{F}^{2}(1+x^{2})}{x-1}\right\} \log x\log (1-x)\nonumber \\
 &  & +\left\{ -\frac{C_{F}}{9}\left[ N_{C}(3\pi ^{2}-67)+20T_{f}\right] \right\} \frac{1}{(1-x)_{+}}\nonumber \\
 &  & +\left\{ C_{F}\left[ \frac{N_{C}(51+44\pi ^{2})-4T_{f}(3+4\pi ^{2})}{72}-3N_{C}\zeta (3)\right. \right. \nonumber \\
 &  & \left. \left. \qquad +C_{F}\left( \frac{3}{8}-\frac{\pi ^{2}}{2}+6\zeta (3)\right) \right] \right\} \delta (1-x)
\end{eqnarray}
\begin{eqnarray}
P^{(1)}_{NS^{-},+-}(x) & = & \left\{ 2C_{F}(2C_{F}-N_{C})(x-1)\right\} \nonumber \\
 &  & +\left\{ C_{F}(N_{C}-2C_{F})(1+x)\right\} \log x\nonumber \\
 &  & +\left\{ \frac{C_{F}(N_{C}-2C_{F})(1+x^{2})}{1+x}\right\} S_{2}(x)
\end{eqnarray}
\begin{equation}
P^{(1)}_{NS^{+},++}(x)=P^{(1)}_{NS^{-},++}(x)
\end{equation}
\begin{equation}
P^{(1)}_{NS^{+},+-}(x)=-P^{(1)}_{NS^{-},+-}(x)
\end{equation}
\begin{eqnarray}
P^{(1)}_{qq,++}(x) & = & \left\{ \frac{C_{F}}{18x}\left[ 2T_{f}(20-(x-1)x(56x-11))\right. \right. \nonumber \\
 &  & \left. \left. \qquad \qquad +x(90C_{F}(x-1)+N_{C}(53-187x+3\pi ^{2}(1+x)))\right] \right\} \nonumber \\
 &  & +\left\{ \frac{C_{F}}{6(x-1)}\left[ 6C_{F}(3-2(x-1)x)-N_{C}(17+5x^{2})\right. \right. \nonumber \\
 &  & \left. \left. \qquad \qquad \qquad +4T_{f}(1+x(x(9+4x)-12))\right] \right\} \log x\nonumber \\
 &  & +\left\{ \frac{C_{F}\left[ C_{F}-N_{C}+4T_{f}-(C_{F}+N_{C}+4T_{F})x^{2}\right] }{2(x-1)}\right\} \log ^{2}x\nonumber \\
 &  & +\left\{ \frac{2C_{F}^{2}(1+x^{2})}{x-1}\right\} \log x\log (1-x)\nonumber \\
 &  & +\left\{ -\frac{C_{F}}{9}\left[ N_{C}(3\pi ^{2}-67)+20T_{f}\right] \right\} \frac{1}{(1-x)_{+}}\nonumber \\
 &  & +\left\{ \frac{C_{F}}{72}\left[ N_{C}(51+44\pi ^{2}-216\zeta (3))-4T_{f}(3+4\pi ^{2})\right. \right. \nonumber \\
 &  & \left. \left. \qquad \qquad +9C_{F}(3-4\pi ^{2}+48\zeta (3))\right] \right\} \delta (1-x)
\end{eqnarray}
\begin{eqnarray}
P^{(1)}_{qq,+-}(x) & = & \left\{ \frac{C_{F}(1-x)}{9x}\left[ 18(2C_{F}-N_{C})x+T_{f}(20-7x+56x^{2})\right] \right\} \nonumber \\
 &  & +\left\{ \frac{C_{F}}{3}\left[ 6C_{F}(1+x)-3N_{C}(1+x)+2T_{f}(3+x(3+4x))\right] \right\} \log x\nonumber \\
 &  & +\left\{ \frac{C_{F}(2C_{F}-N_{C})(1+x^{2})}{1+x}\right\} S_{2}(x)
\end{eqnarray}
\begin{eqnarray}
P^{(1)}_{qg,++}(x) & = & \left\{ \frac{T_{f}}{9x}\left[ N_{C}(20+x(90+x(126+(3\pi ^{2}-218)x)))\right. \right. \nonumber \\
 &  & \left. \left. \qquad -3C_{F}x(12-x(2(15-\pi ^{2})x-3))\right] \right\} \nonumber \\
 &  & +\left\{ \frac{T_{f}}{3}\left[ 6N_{C}+4N_{C}x(12+11x)-3C_{F}(3+2x-4x^{2})\right] \right\} \log x\nonumber \\
 &  & +\left\{ 4T_{f}(C_{F}-N_{C})(1-x^{2})\right\} \log (1-x)\nonumber \\
 &  & +\left\{ T_{f}\left[ 2C_{F}x^{2}-N_{C}(3+x(6+x))\right] \right\} \log ^{2}x\nonumber \\
 &  & +\left\{ 2T_{f}(C_{F}-N_{C})x^{2}\right\} \log ^{2}(1-x)\nonumber \\
 &  & +\left\{ -4C_{F}T_{f}x^{2}\right\} \log x\log (1-x)\nonumber \\
 &  & +\left\{ 2N_{C}T_{f}(1+x)^{2}\right\} S_{2}(x)
\end{eqnarray}
\begin{eqnarray}
P^{(1)}_{qg,+-}(x) & = & \left\{ \frac{T_{f}(x-1)}{9x}\left[ 6C_{F}x(15x-27-\pi ^{2}(x-1))\right. \right. \nonumber \\
 &  & \left. \left. \qquad \qquad \qquad -N_{C}(20-(106+3\pi ^{2}(x-1)-218x)x)\right] \right\} \nonumber \\
 &  & +\left\{ \frac{2}{3}T_{f}\left[ 22N_{C}x^{2}+3C_{F}(3+x(2x-1))\right] \right\} \log x\nonumber \\
 &  & +\left\{ 4T_{f}(N_{C}-C_{F})(1-x)^{2}\right\} \log (1-x)\nonumber \\
 &  & +\left\{ T_{f}\left[ C_{F}(1+2(x-1)x)-N_{C}x^{2}\right] \right\} \log ^{2}x\nonumber \\
 &  & +\left\{ 2T_{f}(C_{F}-N_{C})(1-x)^{2}\right\} \log ^{2}(1-x)\nonumber \\
 &  & +\left\{ -4C_{F}T_{f}(1-x)^{2}\right\} \log x\log (1-x)\nonumber \\
 &  & +\left\{ 2N_{C}T_{f}x^{2}\right\} S_{2}(x)
\end{eqnarray}
\begin{eqnarray}
P^{(1)}_{gq,++}(x) & = & \left\{ \frac{C_{F}}{36x}\left[ 9C_{F}(x-22)x-8T_{f}(10+3x(3x-2))\right. \right. \nonumber \\
 &  & \left. \left. \qquad \qquad +2N_{C}(9-3\pi ^{2}+4x(15+x(18+11x)))\right] \right\} \nonumber \\
 &  & +\left\{ \frac{C_{F}}{3}\left[ 6C_{F}x-N_{C}(12+x(27+4x))\right] \right\} \log x\nonumber \\
 &  & +\left\{ \frac{C_{F}}{3x}\left[ N_{C}(11-3(2-3x)x)-4T_{f}-3C_{F}(3+x(3x-2))\right] \right\} \log (1-x)\nonumber \\
 &  & +\left\{ \frac{C_{F}N_{C}(1+3x(2+x))}{2x}\right\} \log ^{2}x\nonumber \\
 &  & +\left\{ \frac{C_{F}(N_{C}-C_{F})}{x}\right\} \log ^{2}(1-x)\nonumber \\
 &  & +\left\{ -\frac{2C_{F}N_{C}}{x}\right\} \log x\log (1-x)\nonumber \\
 &  & +\left\{ -\frac{C_{F}N_{C}(1+x)^{2}}{x}\right\} S_{2}(x)
\end{eqnarray}
\begin{eqnarray}
P^{(1)}_{gq,+-}(x) & = & \left\{ \frac{C_{F}}{36x}\left[ 27C_{F}(4-5x)x-8T_{f}(10+7(x-2)x)\right. \right. \nonumber \\
 &  & \left. \left. \qquad \qquad +2N_{C}(9-3\pi ^{2}(x-1)^{2}-2x(11-x-22x^{2}))\right] \right\} \nonumber \\
 &  & +\left\{ \frac{C_{F}}{6}\left[ 3C_{F}(4+3x)-8N_{C}(6+(x-3)x)\right] \right\} \log x\nonumber \\
 &  & +\left\{ \frac{C_{F}}{3x}\left[ N_{C}(11+8(x-2)x)-4T_{f}(1-x)^{2}\right. \right. \nonumber \\
 &  & \left. \left. \qquad \qquad -3C_{F}(3+2(x-2)x)\right] \right\} \log (1-x)\nonumber \\
 &  & +\left\{ \frac{C_{F}}{2x}\left[ N_{C}+C_{F}(x-2)x\right] \right\} \log ^{2}x\nonumber \\
 &  & +\left\{ -\frac{C_{F}(C_{F}-N_{C})(x-1)^{2}}{x}\right\} \log ^{2}(1-x)\nonumber \\
 &  & +\left\{ -\frac{2C_{F}N_{C}(x-1)^{2}}{x}\right\} \log x\log (1-x)\nonumber \\
 &  & +\left\{ -\frac{C_{F}N_{c}}{x}\right\} S_{2}(x)
\end{eqnarray}
\begin{eqnarray}
P^{(1)}_{gg,++}(x) & = & \left\{ \frac{1}{18x}\left[ 6C_{F}T_{f}(x-1)(x(37+10x)-2)\right. \right. \nonumber \\
 &  & \left. \qquad \qquad +2C_{F}T_{f}(x(21+x(19+23x))-23)\right. \nonumber \\
 &  & \left. \left. \qquad \qquad +N_{C}^{2}(3\pi ^{2}(x(3+x+x^{2})-1)-x(165+103x))\right] \right\} \nonumber \\
 &  & +\left\{ -\frac{2}{3}\left[ 2N_{C}T_{f}(1+x)+6C_{F}T_{f}(2+x)+N_{C}^{2}(x(14+11x)-1)\right] \right\} \log x\nonumber \\
 &  & +\left\{ \frac{4C_{F}T_{f}x(x^{2}-1)+N_{C}^{2}(1+x(6+x(2+(x-8)x)))}{2(1-x)x}\right\} \log ^{2}x\nonumber \\
 &  & +\left\{ \frac{2N_{C}^{2}(1-x(2-2x-x^{3}))}{(x-1)x}\right\} \log x\log (1-x)\nonumber \\
 &  & +\left\{ -\frac{N_{C}^{2}(1+x+3x^{2}+x^{3})}{x}\right\} S_{2}(x)\nonumber \\
 &  & +\left\{ \frac{N_{C}}{9}\left[ N_{C}(67-3\pi ^{2})-20T_{f}\right] \right\} \frac{1}{(1-x)_{+}}\nonumber \\
 &  & +\left\{ N_{C}^{2}\left( \frac{8}{3}+3\zeta (3)\right) -C_{F}T_{f}-\frac{4N_{C}T_{f}}{3}\right\} \delta (1-x)
\end{eqnarray}
\begin{eqnarray}
P^{(1)}_{gg,+-}(x) & = & \left\{ \frac{1}{18x}\left[ 6C_{F}T_{f}(x-1)(x(7+10x)-2)\right. \right. \nonumber \\
 &  & \left. \qquad \qquad +N_{C}^{2}(2(70-3x)x-3\pi ^{2}(1-x+3x^{2}-x^{3}))\right. \nonumber \\
 &  & \left. \left. \qquad \qquad +2N_{C}T_{f}(x(37+x(23x-57))-23)\right] \right\} \nonumber \\
 &  & +\left\{ 2C_{F}T_{f}(1-3x)-N_{C}^{2}\left( 9-13x+\frac{22x^{2}}{3}\right) \right\} \log x\nonumber \\
 &  & +\left\{ \frac{N_{c}^{2}}{2x}\left( 1-x+3x^{2}-x^{3}\right) \right\} \log ^{2}x\nonumber \\
 &  & +\left\{ \frac{2N_{C}^{2}}{x}\left( x^{3}-3x^{2}+x-1\right) \right\} \log x\log (1-x)\nonumber \\
 &  & +\left\{ -\frac{N_{C}^{2}(1+x(2+2x+x^{3}))}{x(1+x)}\right\} S_{2}(x)
\end{eqnarray}

\end{document}